\documentclass[aps,prl,showpacs,twocolumn,superscriptaddress,groupedaddress]{revtex4}  % for review and submission
\usepackage{graphicx,dblfloatfix}  % needed for figures
\usepackage{dcolumn}   % needed for some tables
\usepackage{bm}        % for math
\usepackage{amssymb}   % for math
\usepackage{amsmath}
\usepackage[english]{babel}
\usepackage[usenames,dvipsnames]{xcolor}

\begin{document}
\title{Dynamics and stability of contractile actomyosin ring in the cell}
\author{Mainak Chatterjee, Arkya Chatterjee$^1$, Amitabha Nandi, and Anirban Sain$^2$\\
\textit{Department of Physics, Indian Institute of Technology-Bombay, Powai, Mumbai 400076, India}  }
\date{\today}

\begin{abstract} Contraction of the cytokinetic ring during cell division leads 
to physical partitioning of a eukaryotic cell into two daughter cells. This 
involves flows of actin filaments and myosin motors in the growing 
membrane interface at the mid-plane of the dividing cell.  Assuming boundary driven alignment of the acto-myosin filaments at the inner edge of the iterface we explore how the resulting active stresses influence  the flow. Using the continuum gel theory framework, we obtain exact axisymmetric solutions of the dynamical equations. These solutions are consistent with experimental  observations on closure rate. Using these solutions we perform linear stability analysis for 
the contracting ring under non-axisymmetric deformations. Our analysis shows 
that few low wave number modes, which are unstable during onset of the constriction, later on become stable when the ring shrinks to smaller radii, 
which is a generic feature of actomyosin ring closure. Our theory also captures how the effective tension in the ring decreases with its radius causing significant slow down in the contraction process at later times.
\end{abstract}

\pacs{87.16.Ka, 87.16.ad, 87.16.dj, 87.17.Ee}
\maketitle
Cell division is fundamental to all living organisms. The last stage of cell division is called cytokinesis, where closure of a polymeric ring, made of actin filaments and myosin molecular motors \cite{myosin2, cytokineticringformation} completes the physical partitioning of the 
cell. %into two daughter cells. 
In one mode of partitioning an intercellular membrane forms (see Fig-1a). This is common in mitotic cell divisions (eg., in {\em C. elegans} embryo, a widely studied model system for eukaryotes) and also in some compact tissues \cite{tissueseptum}. In the other mode (see Fig-1b), the contact area between the daughter cells gradually shrinks to zero, as the division furrow (the cusp in Fig-1b) caves in  \cite{turlier}. Here we focus on the development of the intercellular membrane (the first mode) which starts out as an annulus at the equatorial plane (see Fig-1a and inset-c) and gradually closes itself, as its inner boundary 
{\it grows} radially inward. 
%or embryonic cell division in eukaryotes such as {\em C. elegans}.
%Here we focus on mitotic cell division in {\em C. elegans} embryo, a
%widely studied model system for eukaryotes.
%The growth of the interface is driven by the constriction of 
%a contractile ring which forms at the inner edge of the annular shaped
%interface (see Fig.1-a,c). The ring is made of actin filaments and myosin motors \cite{myosin2, cytokineticringformation}. 
The growth is assisted by the flow of actomyosin, beneath the cell surface ({\it the cortical flow}) 
%, from the polar to the equatorial region of the dividing cell 
\cite{corticalflow}. 
%Experiments have studied the role of actin and myosin in this process \cite{myosin1,myosin2,myosin3}. 
Experiments suggest \cite{myosin1,myosin2,cytokineticringformation} that the ATP driven interaction between actin and myosin lead to the generation of \textit{active} contractile stresses in the cytokinetic ring. How this stress changes with time during the course of the constriction however is not clear.  Earlier models \cite{zumdieck,turlier} explain the observed contraction rate by assuming a constant contractile stress. Ref \cite{anirban} had in addition assumed, an adhoc intrinsic dynamic friction, to account for the 
%on phenomenological grounds, a dynamic component to the stress that causes an 
eventual slowdown of the contraction process.

Such an approach, that considers the actin ring to be a separate entity 
attached with the growing active membrane, cannot explain the recent experimental observations \cite{corticalringlaserablation} where the ring is found to reorganize and constrict even after part of it is destroyed by localized laser ablation. This motivates us to consider the cortical ring to be part of the acto-myosin continuum spread over the growing membrane interface. 
In Ref \cite{ring}, the authors developed an active gel model of the cytoskeletal flows to discuss wound healing in Xenopus oocyte \cite{xenopus}.
%leading to actin ring formation and contraction, in the context of wound healing in Xenopus oocyte \cite{xenopus}. 
Such a description involves solution of coupled equations for the actin alignment field $Q_{\alpha\beta}(r)$ (the order parameter OP) and the velocity field $v_{\alpha}(r)$. The ring was assumed to be a narrow annular zone with higher level of myosin activity $\zeta\Delta \mu$
% and {\color{red}higher degree of actin alignment} 
than the rest of the growing interface. %Continuity of stress and velocity was imposed at the boundary of the ring and the 
%growing cortex. 

In this Letter, we follow a similar continuum gel theory approach and first solve the coupled equations for the OP and the velocity fields numerically (Fig-1), retaining flow coupling. But instead of assuming an active contractility gradient, which is standard in the literature \cite{ring},
%{\color{red}but has not been verified experimentally,} 
we use the observation, that actin filaments are aligned tangentially 
to the inner boundary of the closing annulus \cite{reymann,spira}, as
a boundary condition. This is motivated by recent experiments \cite{align-munro,align-carvalho} which indicate that local assembly kinetics, like guided polymerization, can drive rapid filament alignment at the ring, 
at a much faster rate compared to the relatively slow hydrodynamic modes of the OP and the flow fields. 
The mean (time averaged) effect of this molecular level, fast, alignment kinetics can be incorporated in the hyrodynamic equation for the OP field as a boundary condition.
%therefore does not enter the hydrodynamic equation for the OP. However one can take care ofthe microscopic fast processes by incorporating their mean (time averaged) effect on the slower hydrodynamic modes. 
%than what cortical flows can do.
%This drastically reduces the role of flow coupling as a driver of filament alignment at the boundary (see Fig-1). However the flow coupling still contributes to the spatial tension gradient affecting the velocity field. 
Such a boundary driven alignment was used in Ref\cite{ladu,julicherboundary} to solve for the OP field.  Ref\cite{ladu} reported that acto-myosin filaments at open cell boundaries can respond to the curvature of the boundary, and align parallel or perpendicular to the concave or convex boundaries, respectively. Encouraged by these observations, on boundary driven alignment,
 we set out to compute, a) the constriction rate of the cytokinetic 
 ring, and b) its stability with respect to non-axisymmetric deformations, which has wide applicability across eukaryotic cell division. 
%Our theoretical results from (a) are compared with experimental data on dividing C elegans embryo. Results from
%(b) has wide applicability across eukaryotic cell division. We also
%pin point the roles of curvature, cytoplasmic viscosity, flow coupling and the material inflow at the outer boundary of the annular shaped 
%cell-cell interface.  
%We show when the radius if the hole is large, the circular boundary is unstable to low wave number fluctuations. But as hole decreses in radius, i.e., at later stages of constriction all the modes become stable. 
%by computing the growth rates of the Fourier modes up to reasonably high orders. 

{\it Model: }The actomyosin gel on the growing interface is modeled as a nematic fluid. 
%with  the nematic directors (the filaments) lying on the planar interface. 
Orientational order in a nematic fluid, in 3-dimensions, is defined 
by the tensor order parameter $Q_{\alpha\beta}= \langle n_\alpha n_\beta-\delta_{\alpha\beta}/3\rangle$, where $n_{\alpha}$ is the nematic director field, and $\alpha,\beta=(x,y,z)$. 
%$d$ being the dimension of the embedding space. In this Letter, the embedding dimension being considered is the usual $ d=3 $; hence the Greek subscripts $ \alpha $ and $ \beta $ run over the set $ \{x,y,z\} $. 
As the acto-myosin filaments (nematic directors) lie in the flat interface 
$(x-y)$, symmetry and tracelessness of $Q_{\alpha\beta}$ dictate that the non-diagonal matrix elements involving $z$ are zero, $Q_{xy}=Q_{yx}=q,\; 
Q_{zz}=-1/3$,
and $Q_{xx}+Q_{yy}=1/3$. Further, if the orientation distribution 
is isotropic in the $x-y$ plane then the resulting matrix 
$Q^0_{\alpha\beta}$ is diagonal, with $Q^0_{xx}=Q^0_{yy}=1/6$, and 
$Q^0_{zz}=-1/3$.
%Now, if the orientation of the nematic is isotropically distributed in a 2D ($x-y$) plane (like the actin filaments in the equatorial plane in the absence of cortical flows \cite{isotropicactin1,isotropicactin2}), the nematic order parameter reduces to a diagonal tensor with $ Q _ { x x } ^ { 0 } =  Q _ { y y } ^ { 0 } = 1/6 \text { and } Q _ { z z } ^ { 0 } =- 1/3$ 
%assuming without loss of generality that the 2D plane mentioned above is the XY plane). 
In the presence of cortical flows or due to specific boundary conditions the isotropic distribution is modified to $ Q _ { \alpha \beta } = Q _ { \alpha \beta } ^ { 0 } + Q _ { \alpha \beta } ^ { \prime }$. Again symmetric structure and tracelessness of $Q _ { \alpha \beta }$ require (see Supplementary information -SI) that, $Q^{\prime}_{xx}=-Q^{\prime}_{yy}=\tilde Q$,  
$Q'_{xy}=Q'_{yx}=q$, and rest of the elements are zero. 
This form remains invariant as we transform from cartesian to 2D polar
coordinates later.

\textit{Active gel model for acto-myosin filaments :}
%We study a continuum theory of a 2D active nematic gel expressed in terms of two coarse-grained fields velocity and nematic order parameter. 
The free energy of the inhomogeneous nematic field can be described by the Landau-De Gennes form \cite{degennes}, using the $Q^{\prime}$ matrix. 
%write a Gaussian free energy functional as follows:
%\begin{equation}
$\mathcal {F} = \int d^3r \left(\frac{\chi}{2}Q'_{ij}Q'_{ji}+\frac{L}{2}\partial_kQ'_{ij}\partial_kQ'_{ij}\right)\;.$
%\label{eq.F}
%\end{equation}
This enforces an isotropic arrangement of the director field in the bulk of the 2D 
growing cortical layer with a correlation length $L_c=\sqrt{L/\chi}$. 
Later, we will see that this turns out to be the width of the actomyosin 
ring, which has been measured \cite{oegema} to be $\sim 1\mu m$.
% In writing the above free energy, we have assumed that the nematic is approximately isotropic in the bulk of the two-dimensional cortical layer separating the two halves of the dividing cell.

Constitutive equations of the active gel can be described by a linear relationship between thermodynamic fluxes and forces \cite{ring,activegeltheory1, activegeltheory3,activegeltheory4}. We choose stress tensor $ \sigma_{\alpha\beta} $, the rate of change of nematic order parameter $ \frac{D Q_{\alpha\beta}}{Dt} $, and the rate of ATP consumption as the fluxes. The conjugate forces are the strain rate $ v_{\alpha\beta} =\frac{1}{2}(\partial_\alpha v_\beta+\partial_\beta v_\alpha)$, the traceless nematic force field $H_{\alpha\beta}=-\frac{\delta \mathcal F}{ \delta Q^{\prime}_{\alpha\beta}}$, and the chemical potential difference generated due to ATP hydrolysis $\Delta\mu$. Following
\cite{ring} the hydrodynamic equations in the liquid limit can be expressed as follows:
\begin{eqnarray}
\sigma_{\alpha\beta} &=& 2\eta v_{\alpha\beta} - \beta_1 H_{\alpha\beta} + \zeta\Delta\mu Q_{\alpha\beta}\;, \label{stress}\\
\frac{D}{Dt}Q_{\alpha\beta} &=& \beta_1 v_{\alpha\beta} +  \frac{1}{\beta_2}%H_{\alpha\beta} + \lambda\Delta\mu Q_{\alpha\beta} \label{Q_eqn}
H_{\alpha\beta}\; . \label{Q_eqn}
\end{eqnarray}
$\frac{D}{Dt}$ here implies material derivative \cite{activegeltheory1}, 
$\zeta\Delta\mu Q_{\alpha\beta}$ is the active stress and contractility of the cortical layer enforces $\zeta>0$ \cite{activegeltheory1,activegeltheory3}. We ignored any explicit active term in the second equation because it just renormalizes the inverse susceptibility $\chi^{-1}$. Here $\eta$ is 
the fluid viscosity while $\beta_1$ and  $\beta_2$ are 
Onsager coefficients \cite{ring}, and give the flow coupling and nematic relaxation strengths, respectively \cite{ring}.   

Following \cite{ring}, we define a 2D ``tension tensor" $t_{ij}$ via the relation $t_{ij} = \int (\sigma_{ij} - \delta_{ij}P)dz $. Imposing the 
net normal stress on the interface $t_{zz}$ to be zero yields pressure
$P=\sigma_{zz}$. Further, ignoring variation of stress across the thin interface, we get  \cite{ring} $t_{ij} = e(\sigma_{ij} - \delta_{ij}\sigma_{zz})$, where $e$ is the effective thickness of the interface, assumed to be a constant here.  This tension tensor allows us to write a two-dimensional hydrodynamic theory with the force balance equation as
%\begin{equation}
$\frac{\partial}{\partial t}(\rho v_i) = \partial_j t_{ij} - \alpha v_i$.
%\label{eq.force}
%\end{equation}
Here $ \alpha $ is the cytoplasmic friction external to the growing membrane
interface. The flat growing interface has an annular shape, see  inset 
of Fig.1c. The shrinking cytokinetic ring of radius $R_0(t)$ lies at its 
inner periphery, while its outer periphery is fixed at radius $r_0$.
After changing to 2D polar co-ordinates, and dropping the time derivative 
in highly viscous regime, the force balance equations are 
%$\partial_j t_{ij} = \alpha v_i $. 

%\begin{subequations}
%\begin{align}
%& \partial_r t_{rr} + \frac{1}{r}\left(t_{rr}-t_{\theta\theta}\right) + \frac{1}{r}\partial_{\theta}t_{r\theta} = \alpha v_r\\
%\label{eq.t_rr}
%&\partial_\theta t_{\theta r} + \frac{1}{r}\left(t_{\theta r}+t_{r\theta}\right) + \frac{1}{r}\partial_{\theta}t_{\theta\theta} = \alpha v_\theta\label{eq.trr}
%\end{align}
%\end{subequations}
$\partial_r t_{rr} + \frac{1}{r}\left(t_{rr}-t_{\theta\theta}\right) + \frac{1}{r}\partial_{\theta}t_{r\theta} = \alpha v_r$, and
$ \partial_r t_{\theta r} + \frac{1}{r}\left(t_{\theta r}+t_{r\theta}\right) + \frac{1}{r}\partial_{\theta}t_{\theta\theta} = \alpha v_\theta $
(see SI).

%In the following analysis we make a simplifying assumption mentioned 
%before \cite{alertPRL,alertnatphys}; we ignore flow coupling i.e., $\beta_1\simeq 0$.
 %{\color {red} We transform our equations to 2D polar coordinates in order to take  advantage of the circular symmetry in the $x-y$ plane.}
%and exploit the nearly
%circular structure of the  since the underlying geometry is circular, it is worthwhile to move to 2D polar (or, cylindrical) coordinates 
The $2\times 2\; (xy)$ block of $Q^{\prime}_{\alpha\beta}$ matrix (anisotropic part) remains traceless and symmetric, parameterised by two variables $\tilde Q$ and $q$,
although their values change in the polar frame. 
%depend on the particular coordinate frame used (polar here). 
The $2\times2$ block of the isotropic matrix however remains unchanged, $Q^0_{\alpha\beta}=\mathbb{I}/6$, where $\mathbb{I}$ is the identity matrix (see SI).
% for details).

%Here we have used the zeroth order solutions for $v_r$ and $\tilde Q$ on the right hand sides of the two equations, and the resulting inhomogeneous differential equations are solved using Greens function (see SI).

\textit{Rotationally symmetric solutions for 
$Q^{\prime}_{\alpha\beta}(r)$ and $v_{\alpha}(r)$}: 
We first consider the special case 
%in which we have a (quasi-static) 
where the circular ring is at $r=R_0$, with our domain of interest 
$r\geq R_0$. 
%We neglect external cytoplasmic friction to start with, i.e., 
We start with $\alpha=0$, set stress free boundary condition at 
the open edge, i.e., normal stress $\sigma_{rr}(R_0)=0$, and 
$v_r=0$ at $r\rightarrow\infty$. 
 %The boundary conditions at $r=R_0$ are that the normal component of the stress is zero and 
 The nematic directors are assumed to be parallel to the inner
 boundary, i.e., $\hat n(R_0)=\hat \theta$, %i.e., in the ring. 
 and isotropic as $r\rightarrow\infty$. It implies, that at $r=R_0$, 
 the anisotropic 
 $Q^{\prime}_{\alpha\beta}$ matrix is diagonal with 
 $Q'_{rr}=-Q'_{\theta\theta}=\tilde Q=-1/2$ (see SI), and  $Q^{\prime}_{\alpha\beta}(r=\infty)=0$.
 %We rewrite Eq.\ref{eq.lapl}, for $Q$ and $q$, in their polar form 
%(see SI) : 

We assume a quasi-steady state where the material derivative  $DQ_{\alpha\beta}/Dt=0$ in Eq.\ref{Q_eqn}. Note that $\partial Q_{\alpha\beta}/\partial t\neq 0$ since the the inner edge $R_0$ keeps moving, but the convection term ${\bf v}.\nabla Q_{\alpha\beta}$ counters this change to keep $Q_{\alpha\beta}$ unaltered in the material frame. This yields $H_{\alpha\beta}=-\beta_1\beta_2v_{\alpha\beta}$. When expressed in polar form the 
diagonal elements of this equation gives Eq.\ref{eq.betaQ} below. However the non-diagonal part yields, $q=0$ (see SI). Here we used $\beta_2\approx \eta$ \cite{ring} and 
$\zeta\Delta\mu/\chi\simeq 1$. %Note that $\zeta\Delta\mu/\chi \ll 1$ \cite{ring}, would weaken the flow coupling term on the rhs below.   
\begin{equation}
\frac{1}{r}\partial_r\left(r\partial_r\right)\tilde{Q}-\left(\frac{1}{L_c^2}+\frac{4}{r^2}\right) \tilde{Q} = -\frac{\beta _1}{2L_c^2} \frac{\eta}{\zeta\Delta\mu}\left(\partial_r  v_r - \frac{ v_r}{r}  \right)
\label{eq.betaQ}
\end{equation}

\begin{figure}[h]
	\centering
\includegraphics[width=0.9\linewidth]{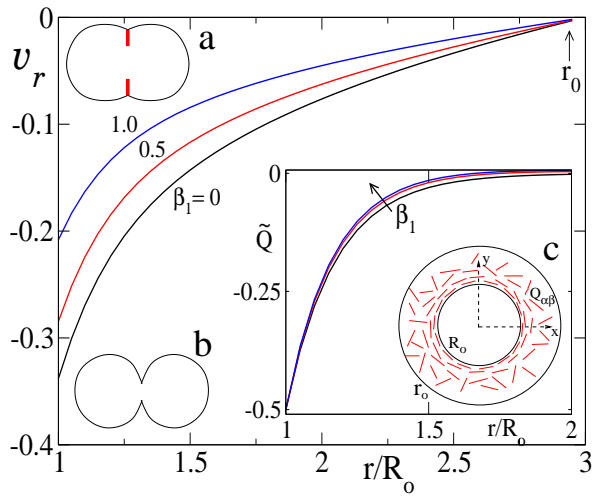}
	\caption{Solutions for the radially symmetric velocity field  $v_r(r)$
(main figure) and the OP field $\tilde{Q}(r)$ (inset) are shown as a function of $r/R_0$, for $R_0=5\mu m$. 
	%(in units of $\frac{\xi \Delta\mu}{\eta}$)
%	 The inset shows	$\tilde{Q}(r)$ as a function of $r/R_0$. 
Schematic diagram 'a'
	shows sideview of the growing interface at the middle of the cell 
	and 'c' shows its cross-sectional view ('b' shows partitioning without an interface).  Alignment of filaments increases sharply near the inner boundary of the annulus at $R_0$. The outer
	boundary is fixed at $r_0=15\mu m$ for these plots. } 
%The lower-right inset shows	the ring radius (scaled by the initial radius, here $R(0)/L_c=10$) as a function of time, in units of $\frac{\eta}{\zeta\Delta\mu}\simeq 3 secs$ here).}
%	, $L_c = 1$ and $\frac{\zeta\Delta\mu}{\eta}=1$}
	\label{fig:v0Rt}
	\end{figure}

Substitution of $H_{\alpha\beta}=-\beta_1\beta_2v_{\alpha\beta}$ into Eq.\ref{stress} simply renormalizes the viscosity to $\tilde\eta = \eta (1 + \frac{1}{2}\beta_1^2)$. The resulting velocity equation (in
polar form) using force balance yields
%(Eq.\ref{eq.betav} below) has the same form as Eq.\ref{Q.eq}.
%Solving the equation below for $\tilde{Q}$,
\begin{equation}
 4\tilde\eta\partial_r\left(\partial_r+\frac{1}{r}\right)v_r= -\zeta\Delta\mu\left(\partial_r+\frac{2}{r}\right)\tilde Q \;, 
\label{eq.betav}
\end{equation}
Using zero influx $v_r(r_0)=0$ at the outer boundary, and a stress 
free inner boundary
$\sigma_{rr}(R_0) = 2\eta \partial_r v_r+\frac{\zeta\Delta\mu}{6}+\zeta\Delta\mu\tilde{Q}=0$,
%$ v_r(r=\infty) = 0$ and,  
%\begin{equation}
%\sigma_{rr}(R_0) = 2\eta \partial_r v_r+
%\frac{\zeta\Delta\mu}{6}+\zeta\Delta\mu\tilde{Q}=0.
%\end{equation}
we solve these two coupled equations numerically (using Mathematica), for different values of the flow coupling strength $\beta_1$, 
%with the same boundary conditions as before (with 
%and at fixed $L=1,\eta=1, \beta_2=1$. 
The solutions are shown in Fig.\ref{fig:v0Rt}, 
%The resulting inward moving velocity field is given below and is plotted in Fig.1, 
using $L_c$ as unit of length and $\frac{\eta}{\zeta\Delta\mu}$ 
as unit of time.
%The resulting solutions include corrections linear in $\beta_1$. 
%In principle the modified $\tilde Q(r)$, gotten from the 1st eqn could be
%used on the rhs of the 2nd eqn, but then we could not get modified $v_r$
%analytically.  
%The boundary conditions used were the same as before. %Probably this was not done, instead the zero-th order solution $\tilde Q^0$ was used in the 2nd equation. 
%Note that, even with nonzero $\beta_1$, the $q=0$ solution holds.
It shows damping of the velocity field $v_r$ with increase in flow coupling strength $\beta_1$. Therefore, stronger flow coupling delays
%can potentially increase 
the ring closure time, however   
%Plots for different coupling parameter $\beta_1$ is shown. 
%While flow coupling damps the flow, 
the order parameter profile, shown in the inset of Fig.\ref{fig:v0Rt}, 
appears to be almost unaffected by flow coupling strength $\beta_1$.
Note that, in this moving boundary problem, the major role of the flow coupling on the OP is to move the boundary inward where the actin field
gets realigned quickly. By setting $\tilde Q(R_0)=-1/2$ we have already
captured this effect indirectly. 
%posteriori justifying our initial assumption of ignoring the flow coupling.
%\begin{equation}
%    v_r^0(r) = -\frac{\zeta\Delta\mu}{\eta}\left[\frac{R_0}{6}\left(1+\frac{3K_1'(R_0/L_c)}{4K_2(R_0/L_c)}\right)\frac{R_0}{r} + \frac{L_c}{8}\frac{K_1(r/L_c)}{K_1(R_0/L_c)}\right] \nonumber
%\end{equation}
This important observation allows us to ignore flow coupling in the
OP equation here (r.h.s. of Eq.\ref{eq.betaQ}) which can now be solved exactly. 
The general solution is $\tilde{Q}(r) = c_1 K_2(r/L_c) + c_2 I_2(r/L_c) $, where $K_2$ and $I_2$ are modified Bessel functions (see SI). 
For outer boundary $r_0\rightarrow \infty$, we get
%$\tilde Q(R_0)=-1/2$ and $\tilde Q(\infty)=0$,  
\begin{equation}
%\tilde{Q}(r) = -\frac{K_2(r/L_c)}{2K_2(R_0/L_c)}\;.
\tilde{Q}(r) = -K_2(r/L_c)/2K_2(R_0/L_c)\;
% \;,\;\; \mbox{and}\;\;\; \label{Q sol symm}%q(r) = 0\,.
 \label{Q sol symm}
%\tilde{Q}^0(r) &=& -\frac{K_2(r/L_c)}{2K_2(R_0/L_c)}\;\;\; \mbox{and}\;\;\; \label{Q sol symm}
\end{equation}
The solution for finite $r_0$ is given in the SI. 
The sharp rise in the magnitude of $\tilde{Q}$ (irrespective of $\beta_1$) at the inner edge can be interpreted as the acto-myosin ring, of width $L_c$. Using this solution we can now solve for $v_r$ (Eq.\ref{eq.betav})
%Eq.\ref{eq.trr} 
with arbitrary $\beta_1$. For $r_0\rightarrow\infty$, the solution reads,
%\begin{equation}
%    4\eta\frac{\partial}{\partial r}\left(\frac{\partial}{\partial r} + %\frac{1}{r}\right)v_r = -\zeta\Delta\mu\left(\frac{\partial}{\partial r}+\frac{2}{r}\right)\tilde{Q}\;,
%\label{Q.eq}
%\end{equation}
%Note that both side of the 2nd equation in Eq.\ref{eq.trr} are 
%identically zero because the non-diagonal components of tension tensor
%as well as $v_{\theta}$ are zero and $t_{\theta\theta}$ is a function 
%of $r$ only.
\begin{eqnarray}
%v_r^0(r) = -\frac{\zeta\Delta\mu}{6\eta}\left[\frac{R_0^2}{r}\left(1-\frac{K_0(R_0/L_c)}{2K_2(R_0/L_c)}\right) + \frac{L_cK_1(r/L_c)}{K_2(R_0/L_c)}\right] \label{zeroth order v}
%%%%v_r(r)= -\frac{\zeta\Delta\mu}{\tilde\eta}
\frac{v_r(r)}{\zeta\Delta\mu/\tilde\eta}= -
\Bigg[\left(1+\frac{3K_1'(R_0/L_c)}{4K_2(R_0/L_c)}\right)\frac{R_0^2}{6r} + \frac{L_c}{8}\frac{K_1(r/L_c)}
{K_2(R_0/L_c)}\Bigg]
\label{zerothorderv}
\end{eqnarray}
%Due to rotational symmetry, of course we have $ v_\theta^0=0 $. 
%Here, the superscript $0$ denotes rotationally symmetric solution. 
Note that the velocity at $r=R_0$, is the ring closure rate
$v_r(R_0)=-\frac{\zeta\Delta\mu}{\tilde\eta} \frac{R_0}{6}\left[1-\frac{3}{4} K_0(R_0/L_c) / K_2(R_0/L_c)\right] $, which is directly damped by the flow
coupling strength $\beta_1$ via the effective viscosity $\tilde \eta$.
%The expression below gives the velocity at $r=R_0$, and is shown in Fig.\ref{zerothorderv}.
%\begin{equation}
%v_r(R_0)=-\frac{\zeta\Delta\mu}{\tilde\eta} \frac{R_0}{6}\left[1-\frac{3K_0(R_0/L_c)}{4K_2(R_0/L_c)}\right]
%\end{equation}
%There are other factor which need to be considered in order to compare these results with experimental data. 
%While these zeroth order solutions reveal the inhomogenous nature of the cortical flow and the filament orientation field, comparisonto experiments requires inclusion of other realistic features. 
Inclusion of cytoplasmic friction ($\alpha \bf v$), 
the velocity influx $v_r (r_0)$ at a finite outer boundary $r=r_0>R_0$ 
(instead of $r_0\rightarrow \infty$) can also influence the flow and 
the closure speed. Solutions for the boundary conditions $\tilde Q(r_0)=0$ 
and $v_r(r_0)=0$ are given in the SI.

%{\it Effect of finite outer radius $r_0$ and cytoplasmic friction :}
%Realistic boundary conditions that both $\tilde Q$ and $v_r$ go to 
%zero at finite outer radius $r_0 >R_0$ (instead of $r_0=\infty$),
%makes the solutions little more complex. The new exact solution for 
%$\tilde Q(r)$, where both $K_2$ and $I_2$ contribute, is given in SI. 
%The corresponding solution for the velocity $v_r$ is obtained 
%numerically using Mathematica.    
%Now we investigate the effect of the cytoplasmic friction which we
%had neglected while arriving at the zeroth order solutions. 

\begin{figure}[h]
	\centering
\includegraphics[width=0.9\linewidth]{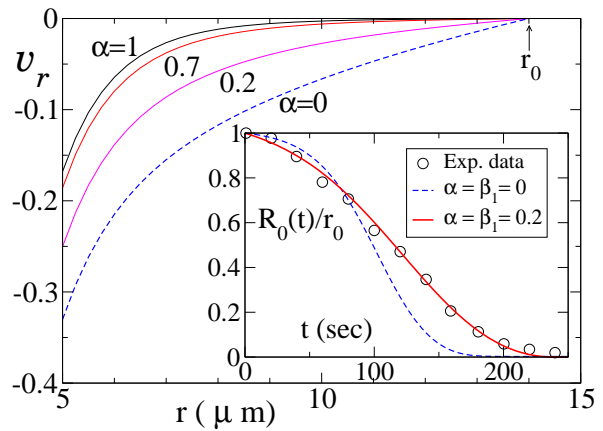}
\caption{Cytoplasmic friction slows down 
the flow: $v_r$ 
%(in units of $\frac{\xi \Delta\mu}{\eta}$)% 
versus $r$ in the main plot for different friction 
coefficients $\alpha$. Inset: lines show scaled    
radius of the ring $R_0(t)/r_0$ versus time (sec),
for $\alpha,\beta_1=0$ and nonzero values (see legends), 
with $\frac{\eta}{\zeta\Delta\mu}=2.06secs$ for both.
%with and without friction and   %The main plot shows 
%$v_r$ as a function of $r$, with 
Furthermore, $r_0=14\mu m,\;$ and $v_r(r_0)=0$. 
%The inset shows that theoretical result 
Circles are the experimental data on C. {\em elegans} 
embryo \cite{zumdieck,oegemaasymmetric}. 
%when cytoplasmic friction and realistic finite outer
%boundary  are incorporated in the theory.  
}  
%at different values of friction constant $\alpha$. 
%The inset shows that the  closure rates scale with initial ring 
%size $r_0$. Here $r/r_0$ are plotted as a function of time, for $r_0=5,10$ 
%and $15\mu m$, and they nearly collapse.}
%The inset shows the corresponding
%ring radius $R_0(t)$, normalised with $R_0(0)$, as a function of time (in
%units of $\frac{\zeta\Delta\mu}{\eta}$).} 
% and initial $R_0=10$}
\label{fig:alpha}
\end{figure}

Cytoplasmic friction adds $\alpha v_r$ to the right hand side 
of Eq.\ref{eq.betav} but does not alter the equation for $\tilde Q$.
Restricting ourselves to radial motion only ($v_r$ nonzero, 
$v_{\theta}=0)$  and assuming azimuthal symmetry, we get
%\begin{equation}
%4\eta\partial_r\left(\partial_r+\frac{1}{r}\right)v_r+\zeta\Delta\mu\left(\partial_r+\frac{2}{r}\right)\tilde{Q} = \alpha v_r \nonumber
%\end{equation}
\begin{equation}
\left[\partial_r\left(\partial_r+\frac{1}{r}\right) - \frac{\alpha}{4\tilde\eta}\right]v_r =-\dfrac{\zeta\Delta\mu}{4\tilde\eta}\left(\partial_r+\dfrac{2}{r}\right)\tilde Q.
%\underbrace{\left[\partial_r\left(\partial_r+\frac{1}{r}\right) - \tilde{\alpha}\right]}_{\mathcal{L}}v_r = f(r)\;,
%\tilde{\alpha } = \dfrac{\alpha}{4\eta} \nonumber
\label{eq.Grn1}
\end{equation}
%where $f(r)=-\dfrac{\zeta\Delta\mu}{4\eta}\left(\partial_r+\dfrac{2}{r}\right)\tilde Q$.  
%We can use $\tilde Q=\tilde Q^0$, since the friction does not change
%Eq.\ref{Q.eq}, while we still ignore flow coupling (i.e., $\beta_1=0$).
 %The zeroth order solution for the order parameter $\tilde Q^0$ remains unchanged as we still ignore flow coupling. 
%Using $\tilde Q=\tilde Q^0$ generates 
%the inhomogeneous right hand side $f(r)$ for the equation for $v_r$. 
%Thus,  = -\dfrac{\zeta\Delta\mu}{4\eta} \dfrac{K_1(r/L_c)}{2K_2(R_0/L_c)}$. 
With boundary conditions $\tilde Q(r_0)=v_r(r_0)=0$, and those at $r=R_0$
remaining same as before, we solve Eq.\ref{eq.Grn1}, both using Green's
function (see SI) and numerically in Mathematica. 
%same boundary conditions at $r=R_0$ and $\tilde Q=0, v_r=0$ at $r=r_0$, 
%on $v_r$ as before,  $v_r(r\to\infty) = 0$, and $\sigma_{rr}(r\to R_0) =0$
%2\eta \partial_r v_r+\frac{\zeta\Delta\mu}{6}+\zeta\Delta\mu\tilde{Q}
%we now solve for $v_r(r)$ which is plotted in Fig.\ref{fig:alpha} for
%different values of $\alpha$. 
As expected, see Fig.\ref{fig:alpha}, cytoplasmic friction damps the 
flow at the growing interface and slows down the ring closure speed 
(inset of Fig.\ref{fig:alpha}). 
%As we already mentioned, the flow is affected by flow-coupling ($\beta_1$), cytoplasmic friction ($\alpha$) and the boundary condition $v_r(r_0)$. Using the closure data alone we cannot estimate all these real parameters. %The nominal fit in the inset is for $\beta_1=v_r(r_0)=0$.

%\textit{ Now finite $r_0$ and $\alpha$.}
The above analysis is carried out quasi-statically for a fixed $R_0$. We can use these results to obtain the ring closure kinetics. We integrate the kinematic boundary condition $\frac{d}{dt}{R_0}=v_r(R_0)$ to derive the time dependence of the radius of the contracting ring i.e., $R_0$ versus $t$. In Fig.\ref{fig:alpha}-inset 
%we make a limited comparison of this closure rate (with $beta_1 =v_r(r_0)=0$) to highlight how much difference $\alpha$ alone can make.
we compare this closure rate  with experimental data on C. {\em elegans} embryo \cite{zumdieck,oegemaasymmetric}. Note that this is a three parameter fit with $\alpha, \beta_1$ and the active time scale $\frac{\eta}{\zeta\Delta\mu}$. Reasonable fits can be obtained for several 
combinations of these parameters in the range $\alpha,\beta_1\in 
[0.1,0.5]$ and $\frac{\eta}{\zeta\Delta\mu}\in [1.5,2.5] sec$. One such example 
is shown in Fig.\ref{fig:alpha}-inset. Here we used $L_c=1 \mu m$ \cite{oegemaasymmetric}. 
%in the range  fit with $\alpha, The data can be fitted We use $L_c$ ($1 \mu m$ \cite{oegemaasymmetric}) as unit of length and $\frac{\eta}{\zeta\Delta\mu}$ as unit of time.
%Choosing model parameters $L_c=1, \frac{\zeta\Delta\mu}{\eta}=1$ in eq \eqref{zeroth order v} (a choice that is equivalent to adopting $ \dfrac{\zeta\Delta\mu}{\eta} $ as the unit of inverse time and $ L_c $ as the unit of length.), 
%The theoretical result fits the experimental data well and 
%Using $\frac{\eta}{\zeta\Delta\mu}\simeq 2.15$ sec
%and the friction constant $\alpha\simeq 0.7$. 
Membrane tension $\sigma_0$ in the growing membrane can be linked to the activity as $\sigma_0=\zeta\Delta\mu e/2$ \cite{ring}. Using $\frac{\eta}{\zeta\Delta\mu}\simeq 2$ secs, measured value of cortical tension 
 $\sigma_0= 3\times10^ {-4}N/m$ \cite{memtension} and the thickness of the growing actomyosin cortex $e\simeq 0.3\mu m$ \cite{oegemaasymmetric}, we get 
$\eta\simeq 4\times 10^3$ Pa.sec, which is similar to the estimates 
obtained in earlier works \cite{ring,viscosity}.

\begin{figure}[h]
	\centering
	\includegraphics[width=1.0\linewidth]{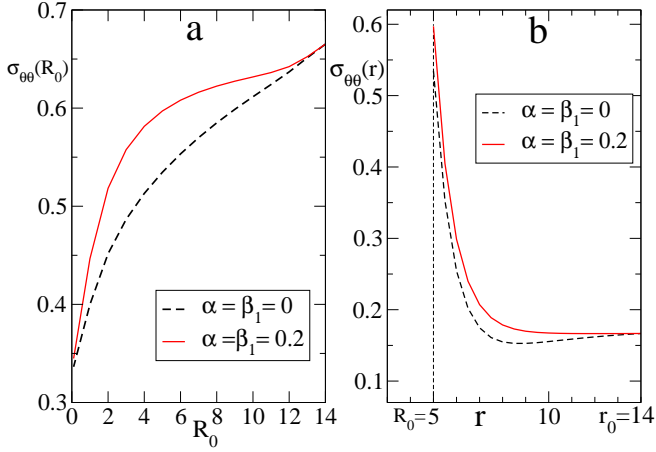}
	\caption{The line tension, (Eq.\ref{eq.tt}), at the ring 
	$\Sigma=\sigma_{\theta\theta}(R_0)$,
	in units of activity $\zeta\Delta\mu$, shown
	as a function of ring radius $R_0$ (in a), and as a function of $r$ 
	(in b), 
	at a fixed $R_0=5\mu m$.  The outer radius is fixed at $r_0=14\mu m$,
	appropriate for C. {\em elegans} embryo \cite{zumdieck, oegemaasymmetric}. Friction (nonzero $\alpha$) does not affect ring tension
	significantly.
	} 
	\label{fig:sigmatt}
\end{figure}

%{\it Slowing down of ring closure rate : } 
The ring closure rate in eukaryotes shows an intriguing 
slow down at late times (Fig\ref{fig:alpha}-inset), which has not been understood yet. 
%Reduction in the effective ring tension or increase in internal friction have been proposed as possible reasons. 
In Ref \cite{anirban} an adhoc intrinsic dynamic friction 
$\zeta_L$ was added,  to the ring tension 
%produced by the ring. This adhoc term was needed on phenomenological grounds 
to account for hitherto unknown internal processes in the ring.
In Ref \cite{turlier} the cortical flow from the poles, converging towards the equatorial furrow ($v_r(r_0)$ in our theory), was shown to affect the slow down \cite{turlier}. 
% {\color{red} although no comparison to real data was made: check}. 
%In our theory it is represented by the influx $v_r(r_0)$ at the outer boundary of the inward growing interface. We have assumed $v_r(r_0)=0$ so far, however using a nonzero value, instead, would make the closure faster. 
% which might be responsible for the slow down.
In our present theory $\sigma_{\theta\theta}(r=R_0)$ is the effective ring tension $\Sigma$ of Ref \cite{anirban}. From Eq.\ref{stress},
\begin{equation}
\sigma_{\theta\theta} = 2\eta\frac{v_r}{r}  + \frac{\zeta\Delta\mu}{6} - \zeta\Delta\mu \tilde{Q}. 
\label{eq.tt}
\end{equation} 
%This is plotted in Fig.\ref{fig:sigmatt}. 
In Fig.\ref{fig:sigmatt}a we show
the ring tension as a function of the ring size $R_0$, and 
Fig.\ref{fig:sigmatt}b shows how azimuthal stress varies in the 
bulk of the closing interface,
%$\sigma_{\theta\theta}$ over the whole membrane 
for a given ring size $R_0$. First, $\sigma_{\theta\theta}$ is always
positive, implying contractile stress in the ring and the interface. Second, the ring tension falls sharply at small $R_0$, which explains the slow down. Third, the azimuthal stress $\sigma_{\theta\theta}(r)$ is very high at the edge $r=R_0$ and small in the interior. This property perfectly justifies the role of the ring as the main generator of cytokinetic tension.
Note, that in Eq.\ref{eq.tt} the last two terms on the right hand side are constants (at the ring $r=R_0$), and positive. But the first term is negative and its magnitude grows large as the hole shrinks, eventually reducing the line tension.    
So the slowing down effect appears naturally due to viscosity of the flowing gel and curvature of the ring. Interestingly this tension reduction term has  the same structure $v_r(R_0)/R_0=\dot R_0/R_0$ which was assumed in Ref \cite{anirban}, based purely on phenomenology.

\begin{figure}[h]
	\centering
	\includegraphics[width=0.95\linewidth]{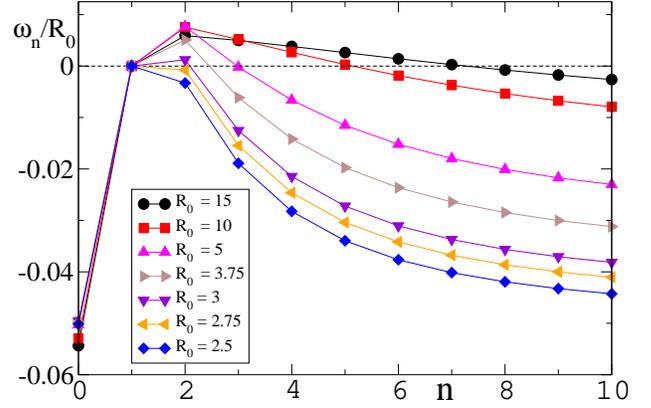}
	\caption{$\omega_n$, scaled by the inner radius $R_0$, as a
	function of the mode number $n$ for different values of  $R_0$ ($\mu m$), see legends. Low wave number modes switches from unstable to 
	stable at smaller $R_0$. }	
	\label{fig:omega_n_5}
\end{figure}

{\it Stability of ring closure : }
We now use the rotationally symmetric solutions for the $Q_{ij}$ and the
${\bf v}$ fields to examine the stability of the inner boundary where
the ring forms. This is motivated by the observation that wild type 
rings, during constriction, typically show deviation from circular 
shape \cite{oegemaasymmetric, vidya, corticalringlaserablation}, however it becomes more circular as constriction proceeds.
%Also   rconstriction shows that typically the ring is a deformed This is motivated by the observation that most often the ring constrictsas a deformed circle and only near the end starts out as a deformed circle and by Silva et al.\cite{corticalringlaserablation}, as mentioned earlier. Fig.\ref{fig:laser_expt} from their paper\cite{corticalringlaserablation} shows how after a localised laser ablation of the actin ring it heals and completes the closure in almost the sametime as an unperturbed wild type actin ring. 
%Also we note that most wild type rings are not rotationally symmetric (i.e., non-circular) in the beginning of the cytokinesis but evolves into circular shape as the closure progresses \cite{oegemaasymmetric}.
%While weak deviation from circularity can be quantified by studying the dynamics of the lowest few fourier modes, a laser cut involves generation of the relatively high fourier modes. 
%We will show all these modes are stable implying the robustness of the contraction process. 
%We include an image from their paper below (\textbf{note} that this image is not the authors' work). 
Towards this we express the shape of the deformed inner boundary, at any given 
time, as $r(\theta)=R_0 +\delta R(\theta)$, and using Fourier decomposition
%We study deformations of the circular ring, decomposed into Fourier modes as follows:
%\begin{equation}
$\delta R(\theta,t)=\sum_{n=0}^{\infty} \delta R_n e^{in\theta+\omega_n t}\;.$
%\end{equation}
We study stability of these deformation modes \cite{alertnatphys} by computing $\omega_n$, 
up to $n=10$. Note that the $n=1$ mode corresponds to an uniform translation of the inner circular boundary and therefore $\omega_1= 0$. The system has translational symmetry provided the outer boundary $r_0\rightarrow \infty$, which we exploit for this calculation. The results below are unlikely to change qualitativey when $r_0$ is finite, except that $\omega_1$ will be nonzero.
%This perturbative deformation to the ring implies that the circular ring defined by the polar equation $r(\theta)=R_0$ is now changed to $r(\theta)=R_0+\delta R(\theta)$ (the time dependence is suppressed for simplicity of notation). 

%\begin{figure}[h]
%	\centering
%	\includegraphics[width=0.6\linewidth]{deformed_ring.png}
%	\caption{Deformed hole}
%	\label{fig:deformed_ring}
%\end{figure}
The change at the inner edge leads to change in all the dynamical variables : 
%Velocity and Order parameter fields change perturbatively:
%   \begin{eqnarray}
        $\tilde{Q}(r,\theta,t) =\tilde{Q}_0(r)+\delta\tilde{Q}(r,\theta,t)$, 
        and similarly,
        $q (r,\theta,t) = \delta q(r,\theta,t),\;
        v_r(r,\theta,t) = v_r^0(r)+\delta v_r(r,\theta,t),\;$ and 
        $v_\theta(r,\theta,t) = \delta v_\theta(r,\theta,t).$ 
%    \end{eqnarray}
  
Further, the perturbation fields $\delta \tilde{Q}, \delta q, \delta v_r$, and $\delta v_{\theta}$ can be decomposed into Fourier modes as  
%$\delta R(\theta,t)$ as follows:}
$\delta\tilde{Q}(r,\theta,t) = \sum_{n=0}^{\infty} \delta \tilde{Q}_n(r) e^{in\theta+\omega_n t}$, $\;\delta v_r(r,\theta,t) = \sum_{n=0}^{\infty} \delta v_{r,n}(r) e^{in\theta +\omega_n t}$, and similarly for the other two fields.

%\begin{eqnarray}
%    \delta\tilde{Q}(r,\theta,t) &=& \sum_{n=0}^{\infty} \delta \tilde{Q}_n(r,t) e^{in\theta} \nonumber \\
%    \delta q(r,\theta,t) &=& \sum_{n=0}^{\infty} \delta q_n(r,t) e^{in\theta} \nonumber\\
%    \delta v_r(r,\theta,t) &=& \sum_{n=0}^{\infty} \delta v_{r,n}(r,t) e^{in\theta} \nonumber \\
%    \delta v_\theta(r,\theta,t) &=& \sum_{n=0}^{\infty} \delta v_{\theta,n}(r,t) e^{in\theta} 
%\end{eqnarray}
We substitute these perturbed fields in the dynamical equations and do a
linear stability analysis to obtain $\{\omega_n\}$, where $\omega_n=\partial_r v_r^0(R_0) + \frac{\delta v_{r,n}(R_0)}{\delta R_n}$, following Ref \cite{alertnatphys}. Details of our calculations are given in the SI.
%supplementary material \cite{SI}.
%We do a linear stability analysis of these deformations to obtain the growth rate $ \omega_n $ of each Fourier mode \cite{SI}.
%The perturbation fields $\delta \tilde{Q}, \delta q, \delta v_{r,\theta}$ can be expanded similar to $\delta R(\theta,t)$. For example, $ \delta\tilde{Q} $ is expanded as:
%\begin{equation}
%\delta\tilde{Q}(r,\theta,t) = \sum_{n=0}^{\infty} \delta \tilde{Q}_n(r,t) e^{in\theta}
%\end{equation}
%We substitute these expansions in the order parameter dynamics and use the exact forms of the zeroth order (rotationally symmetric) solutions derived earlier to obtain expressions for the Fourier amplitudes of all the 4 fields. 
%\begin{figure*}[!b]
%	\centering
%	\includegraphics[width=\textwidth]{fourier_modes.png}
%	\caption{Various Fourier modes with undeformed ring in dotted lines}
%	\label{fig:fourier_modes}
%\end{figure*}

%These solutions can be used in the hydrodynamic differential equations to solve for $ \delta v_r $ and $ \delta v_\theta $ \cite{SI}.
Fig.\ref{fig:omega_n_5} reveals interesting behaviour for the growth 
rates of the Fourier modes $\{\omega_n\}$ for different inner radius
$R_0$. At large $R_0$ several modes are unstable ($\omega_n>0$), however they subsequently turn stable ($\omega_n<0$) as $R_0$ becomes small, 
absolutely consistent with experimental observations. Note that 
$\omega_0 <0$, irrespective of $R_0$, implies stability with respect 
to uniform contraction or expansion of the circular inner boundary.
While in our theory $\omega_n$ is exactly proportional to the activity, 
Fig.\ref{fig:omega_n_5} shows that $\omega_0$ is approximately 
proportional to $R_0$. Also note that the higher modes decay relatively 
faster which would make any sharp distortion of the ring heal fast. This
could be relevant for would healing in cells as well. But the fact 
that larger number of modes are unstable at larger ring size indicates 
that very large rings, if distorted, will fail to contract. 
%We also checked that activity controls  the magnitude of $\omega_n$ but not its sign.

%up to order $n=10$ are plotted in fig \ref{fig:omega_n_5}. All the $\{\omega_n\}$ turn out to be negative, except $n=1$ which can be shown to be zero. The $n=1$ mode corresponds to net shift of the center of mass of the hole which should not excite any dynamics in the system due to translational invariance. While negative values of $\omega_n$ implies that the corresponding Fourier deformation mode is stable to perturbation, $\omega_1=$ implies marginal stability of the shifted 
%mode $n=1$.  But contrary to our expectation that higher (small wave-length) modes would die out relatively faster compared to large wave length modes, $\omega_n$ tends
%to saturate at higher $n$ values.   For generating this plot, we have chosen $ L_c $ as our unit of length and $ \frac{\eta}{\zeta\Delta\mu} $ as our unit of time. Further, we have chosen $ R_0/L_c=5 $.  
%symmetry arguments say that $n=1$ mode must be marginally stable, i.e. $\omega_1=0$ \cite{SI}.

%{\bf Discussion} 
In summary, our phenomenological approximation on the boundary driven actomyosin alignment, was useful in obtaining exact solutions for the OP and the velocity field.  The stability calculation, which produced 
several insights, exploited these solutions to perturb around them. 
%In summary, our treatment of the ring as a part of the acto-myosin continuum, and the phenomenological boundary condition of perfect allignment at the inner edge of the annulus led to analytic expressions for the OP and the velocity field. The stability calculations of the deformed ring, which offered several insights, would not have been possible without these analytic results. 
Also we could identify three separate sources of slow down near the end of the contrictions, namely, a) the curvature at the ring ($1/R_0$), b) the cytoplasmic friction ($\alpha$),  and c) the flow coupling strength 
($\beta_1$). Experiments along the lines of Ref\cite{align-munro}
which probed poly/depolymerization processes near the ring and Ref\cite {corticalringlaserablation} which studied healing of the perturbed 
ring after laser ablation, might be useful to assess the role of 
of boundary in maintaining actin alignment in the dynamic ring.  
%This effect could be relevant for wound healing in tissues as well.
%For example,  wound healing in Drosophila wing imaginal discs shows
%\cite{tissue} a similar slow down which is attributed \cite{tissue} 
%to weakening of the actin purse-string, an inter-cellular structure 
%similar to intra-cellular actin ring.

Acknowledgement: 
We thank one of the referees for pointing out Ref\cite{ladu} to us. 
MC would like to thank IIT Bombay, India for financial support. AN and 
AS acknowledge Science and Engineering Research Board (SERB), India
Project No. ECR/2016/001967 and CRG/2019/005944, respectively, for 
financial support. MC and AC would like to thank Dr. R. Alert for sharing his stability 
calculations in Ref \cite{alertnatphys}.   

\noindent $^1$ Present address: Physics Department, MIT, Cambridge MA 02139, USA.\\
$^2$ asain@phy.iitb.ac.in,   The 1st and the 2nd authors have contributed equally to this work.

\bibliography{main1}
\end{document}

\title{ {Supplementary Information : Part-I}\\
Dynamics and stability of the contractile actomyosin ring in the cell}
%of the cytokinetic ring in weak flow coupling limit.}
%On the Stability of Cytokinetic Contractile Ring Constriction}
%\input author_list.tex
\author{ Mainak Chatterjee, Arkya Chatterjee, Amitabha Nandi, and Anirban Sain\\
\textit{Department of Physics, Indian Institute of Technology Bombay, Powai, Mumbai 400076, India}}  
\maketitle
%\date{\today}

\section {1. Structure of the order parameter matrix
$Q_{\alpha\beta}= n_{\alpha} n_{\beta}-\delta_{\alpha\beta}/3$}
\label{op}
\subsection {1.1. $Q_{\alpha\beta}$ for filaments lying in the $x-y$ plane}
When the filaments lie entirely in the $x-y$ plane, i.e., $\hat n= (n_x,n_y,0)$ then the symmetric, traceless matrix,

\begin{align}
\mathbb{Q} 
= 
\begin{bmatrix} 
Q_1&q&0\\
q&Q_2&0\\
0&0&-1/3
\end{bmatrix}\notag
%\\
%&=
\;,\mbox{ with}\;\; Q_1+Q_2=1/3
\end{align}

When the distribution of filaments is random (isotropic) in the X-Y
plane then due to $\langle n_x^2\rangle=\langle n_y^2\rangle=1/2$ 
and $n_z=0$, the resulting $\mathbb{Q}^0$ is diagonal, with 
$Q_{xx}=Q_{yy}=1/6$ and $Q_{zz}=-1/3$. 
Due to flow or boundary condition $\mathbb{Q}$ will deviate from
its isotropic form and thus,  $\mathbb{Q}=\mathbb{Q}^0 + \mathbb{Q}^{\prime}$. Explicitly,
\begin{align}
 \begin{bmatrix} 
Q_1&q&0\\
q&Q_2&0\\
0&0&-1/3
\end{bmatrix}\notag
&= 
\begin{bmatrix} 
1/6&0&0\\
0&1/6&0\\
0&0&-1/3
\end{bmatrix}\notag
+
 \begin{bmatrix} 
q_1&q&0\\
q&q_2&0\\
0&0&0
\end{bmatrix}\notag 
,\mbox{ preserving symmetric structure of the L.H.S.}\; .
\end{align}
Further, $\mathbb{Q}^{\prime}$ also must be traceless, since
 $\mathbb{Q}$ and  
$\mathbb{Q}^0$ are already tracelsss, and thus $q_1=-q_2\;(=\tilde Q$ in
the main text).

\begin{figure}[h]
	\centering
	\includegraphics[width=0.35\linewidth]{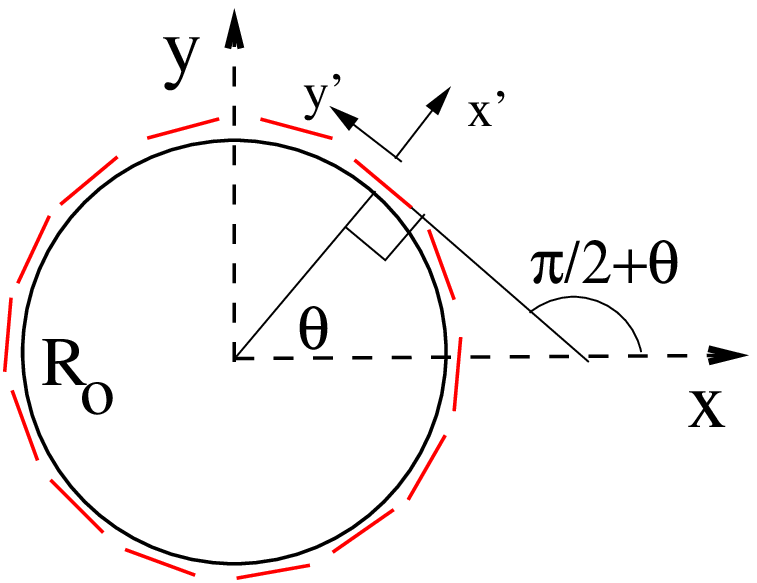}
	\caption{Nematic aligned along the azimuthal direction}
	\label{fig.sup_bc}
\end{figure}

\subsection{1.2. Boundary condition for the order parameter $Q_{\alpha\beta}$}
We impose that the acto-myosin filaments (red lines in Fig.\ref{fig.sup_bc} below) be completely aligned with the 
circular inner boundary, at $r=R_0$, i.e., they are parallel to the azimuthal direction $\hat \theta$. At the outer boundary, either at a finite radius $r_0 >R_0)$ or at $r\rightarrow \infty$, the filaments are randomly oriented. This is a phenomenological input to our theory, ensuring that an acto-myosin ring forms at the open edge of the inward growing membrane. Below, we derive the corresponding order parameter tensor in the cylindrical polar coordinates.

Let the nematic director on the ring, at an angle $\theta$,
be oriented along $\hat n (\theta)$, (see Fig.\ref{fig.sup_bc}). In the Cartesian reference frame $(x,y)$,  $\hat n (\theta)=(n_x,n_y,0)=\hat \theta=(-\sin{\theta},\cos{\theta},0)$. The resulting order parameter matrix at $r=R_0$ is then
\begin{align}
\mathbb{Q}(R_0,\theta) 
&= 
\begin{bmatrix} 
n_x^2-\frac{1}{3}&n_x n_y &0\\
n_x n_y&n_y^2-\frac{1}{3}& 0\\
0&0&-1/3
\end{bmatrix}\notag
=
\begin{bmatrix} 
\sin^2\theta-\frac{1}{3}&-\frac{\sin{2\theta}}{2}&0\\ 
-\frac{\sin{2\theta}}{2}& \cos^2\theta- \frac{1}{3}&0\\
0&0&- \frac{1}{3}
\end{bmatrix}
\end{align}

But $\mathbb{Q}$ has a simple form in the polar coordinate 
frame ($\hat r,\hat\theta$), which is a obtained from the 
original cartesian frame $(x,y)$ via a counter-clockwise 
rotation (by angle $\theta$). In the polar frame, 
$\hat n (\theta) = (0,1,0)$, and thus,
\begin{align}
\mathbb{Q}^p(R_0,\theta)=
\begin{bmatrix} 
-1/3&0&0\\
0&2/3&0\\
0&0&-1/3
\end{bmatrix}\notag.
\end{align}
Here the superscript $p$ denotes polar.
We seek the same decomposition of $\mathbb{Q}^p$ into 
isotropic and deviation parts:
$\mathbb{Q}^p=\mathbb{Q}^{0p} + \mathbb{Q}^{\prime p}$.
However $\mathbb{Q}^{0p}$ remains same in the two coordinate
frames. Thus,
\begin{align}
\mathbb{Q}^{\prime p}(R_0,\theta)=
\begin{bmatrix} 
-1/3&0&0\\
0&2/3&0\\
0&0&-1/3
\end{bmatrix}
- 
\begin{bmatrix} 
1/6&0&0\\
0&1/6&0\\
0&0&-1/3
\end{bmatrix}
=
\begin{bmatrix} 
-1/2&0&0\\
0&1/2&0\\
0&0&0
\end{bmatrix}.
\end{align}
This serves as the boundary condition at $r=R_0$ in our radially
symetric formulation. Here we obtained the polar forms of the 
$\mathbb{Q}$ matrices using the transformed form of director 
$\hat n$ in the polar frame. One could also transform the 
$\mathbb{Q}$ matrix from the cartesian frame $(x,y)$ to the
polar frame ($\hat r,\hat\theta $), using $\mathbb{Q}^p=R\mathbb{Q}R^{-1}$, where $R$ is the rotation matrix for counter-clockwise
passive rotation about $\hat z$ axis, by angle $\theta$.
%Transforming this matrix to the polar coordinate frame ($\hat r,\hat\theta$) fixed at $\theta$, amounts to rewriting $\mathbb Q$ in a rotated coordinate system. The rotated frame ($\hat r,\hat\theta$) is obtained from $(x,y)$ by an anti-clockwise rotation
%through angle $\theta$. If the rotation matrix is $R$ then the order-parameter matrix will transform as $Q=R\mathbb{Q}R^{-1}$. 
More explicitly, 
\begin{align}
%\begin{bmatrix}
%Q_{rr}&Q_{r\theta}\\
%Q_{\theta r}&Q_{\theta\theta}
%\end{bmatrix}
%\\
\mathbb{Q}^p
&= \begin{bmatrix}
\cos{\theta}&\sin{\theta}&0\\
-\sin{\theta}&\cos{\theta}&0\\
0&0&1
\end{bmatrix}
\begin{bmatrix} 
\sin^2\theta-\frac{1}{3}&-\frac{\sin{2\theta}}{2}&0\\ 
-\frac{\sin{2\theta}}{2}& \cos^2\theta- \frac{1}{3}&0\\
0&0&- \frac{1}{3}
\end{bmatrix}
%\begin{bmatrix} 
%-\frac{\cos{2\theta}}{2}&-\frac{\sin{2\theta}}{2}\\ -\frac{\sin{2\theta}}{2}&\frac{\cos{2\theta}}{2}
%\end{bmatrix}
\begin{bmatrix}
\cos{\theta}&-\sin{\theta}&0\\
\sin{\theta}&\cos{\theta}&0\\
0&0&1
\end{bmatrix}
= 
\begin{bmatrix} 
-1/3&0&0\\
0&2/3&0\\
0&0&-1/3
\end{bmatrix}
\end{align}
Using the same transformation on $\mathbb{Q}^0$ one can show that the isotropic part remain unchanged in the polar frame.

\section{2. Constitutive relations and the force-balance equation in 2D polar coordinates}%: Weak flow coupling limit}
\label{ces}
Recall the constitutive equations (Eq.~(2),(3) in the maintext) for the stress and the nematic order parameter
\beqr
\label{ce_1}
\sigma_{\alpha\beta} &=& 2\eta v_{\alpha\beta}-\beta_1H_{\alpha\beta}+\zeta\Delta\mu Q_{\alpha\beta}, \\
\frac{D}{Dt}Q_{\alpha\beta}&=& \beta_1 v_{\alpha\beta}+\frac{1}{\beta_2} H_{\alpha\beta}.
\label{ce_2}
\eqnr

In the steady state $H_{\alpha\beta}=-\beta_1 \beta_2 v_{\alpha\beta}$. Substitution of  $H_{\alpha\beta}$ into Eq.3 renormalises the viscosity to $\bar \eta = \eta (1+\frac{\beta_1^2\beta_2}{2\eta})$. Further, using the definition $Q_{\alpha\beta}=Q^0_{\alpha\beta}+Q'_{\alpha\beta}$ and retaining the same notation $\eta$ (henceforth) for the renormalised viscosity, the components of the stress tensor given in Eq.~\ref{ce_1} can be written in 2D polar coordinates $(r,\theta)$ as:
\beqr
\label{stress_p1}
\sigma_{rr} &=&2\eta \partial_r v_r+\frac{\zeta\Delta\mu}{6}+\zeta\Delta\mu\tilde{Q},\\
\sigma_{r\theta} &=& \eta\left(\frac{1}{r}\partial_\theta v_r+\partial_r v_\theta - \frac{v_\theta}{r}\right) + \zeta\Delta\mu q,
\\
\sigma_{\theta r} &=& \sigma_{r\theta},\\
\sigma_{\theta\theta} &=& 2\frac{\eta}{r} \left(\partial_\theta v_\theta + v_r\right) + \frac{\zeta\Delta\mu}{6} - \zeta\Delta\mu \tilde{Q}.
\label{stress_p2}
\eqnr
Here $\sigma_{rz}=\sigma_{zr}=\sigma_{\theta z}=\sigma_{z \theta}=0$. The components of the strain rate tensor $v_{\alpha\beta}$, in polar coordinates, are
%The strain rates $v_{\alpha\beta}$ in 2D polar coordinates read
\begin{equation}
v_{\alpha\beta} = \begin{bmatrix} \partial_rv_r & \frac{1}{2}\left\{\frac{1}{r}\partial_{\theta}v_r + r\partial_r\left(\frac{v_{\theta}}{r}\right)\right\} \\ \frac{1}{2}\left\{\frac{1}{r}\partial_{\theta}v_r + r\partial_r\left(\frac{v_{\theta}}{r}\right)\right\} & \frac{1}{r}(v_r+\partial_{\theta}v_{\theta}) 
\end{bmatrix}
\label{eq.Qq}
\end{equation}

Furthermore we obtain $\sigma_{zz}$ using the traceless condition.
\begin{equation}
\label{stress_z}
\sigma_{zz} = -(\sigma_{rr}+\sigma_{\theta\theta}) = -2\eta\left(\frac{1}{r}v_r+\partial_rv_r+\frac{1}{r}\partial_\theta v_\theta\right) - \frac{\zeta\Delta\mu}{3}
\end{equation}
%\red{The 2D tension tensor for the cortical layer is given by $t_{ij}=\int{dz(\sigma_{ij}-P\delta_{ij})}$ \cite{salbreux}. Here $P$ is the pressure on the cortical layer which can be obtained by demanding $t_{zz}=0$ \cite{salbreux} . This gives $P=\sigma_{zz}$ and the 2D tension tensor can now be written as $t_{ij}=e(\sigma_{ij}-\sigma_{zz}\delta_{ij})$. Here $e$ is the thickness of the actomyosin layer in which the ring is being formed, and is assumed to be a constant. Using this definition of $t_{ij}$, the force-balance equation $\partial_jt_{ij}=\alpha v_i$ can be written in polar coordinates as (Eq.~7 in the maintext) as}
Now we define the effective 2D tension tensor for the cortical layer $t_{ij}=\int{dz(\sigma_{ij}-P\delta_{ij})}$ 
\cite{salbreux}. As explained in maintext (originally from  
Ref\cite{salbreux}),  $P=\sigma_{zz}$. The corresponding force balance equations, $\partial _j t_{ij}=\alpha v_i$, in polar cordinates read
\beqr
\label{fb1}
&&\partial_r t_{rr} + \frac{1}{r}\left(t_{rr}-t_{\theta\theta}\right) + \frac{1}{r}\partial_{\theta}t_{r\theta} = \alpha v_r, \\
&&{\partial_r}  t_{\theta r} + \frac{1}{r}\left(t_{\theta r}+t_{r\theta}\right) + \frac{1}{r}\partial_{\theta}t_{\theta\theta} = \alpha v_\theta\label{eq.trr}.
\label{fb2}
\eqnr
Substituting Eqs.~(\ref{stress_p1}--\ref{stress_z}) and Eq.6 from maintext into Eqs.~(\ref{fb1}--\ref{fb2}), we get the full expression of the force-balance equation

\beqr
\label{fb_full1}
&&2\eta\left(2\frac{\partial}{\partial r}\left(\frac{\partial}{\partial r} + \frac{1}{r}\right)v_r + \frac{3}{2}\frac{\partial}{\partial r}\left(\frac{\partial}{\partial \theta}\frac{v_\theta}{r}\right)+\frac{\partial}{\partial \theta}\left(\frac{\partial}{\partial \theta}\frac{v_r}{2r^2}\right)-\frac{\partial}{\partial \theta}\frac{v_\theta}{r^2}  \right)+\zeta\Delta\mu\left(\left(\frac{\partial}{\partial r}+\frac{2}{r}\right)\tilde{Q}+\frac{\partial}{\partial \theta}\frac{q}{r}\right) = \alpha  v_r,  \\ 
&& \eta\left(\frac{\partial}{\partial r}\left(\frac{\partial}{\partial r}v_\theta\right)+ 3\frac{\partial}{\partial r}\left(\frac{\partial}{\partial \theta}\frac{v_r}{r}\right)+5\frac{\partial}{\partial \theta}\frac{v_r}{r^2}+4\frac{\partial}{\partial \theta}\left(\frac{\partial}{\partial \theta}\frac{v_\theta}{r^2}\right)-\frac{v_\theta}{r^2} +{\frac{1}{r}}\frac{\partial}{\partial r}v_\theta  \right) + \zeta\Delta\mu\left(\left(\frac{\partial}{\partial r}+\frac{2}{r}\right)q-\frac{\partial}{\partial \theta}\frac{\tilde{Q}}{r}\right) = \alpha  v_\theta. \nonumber \\
&& 
\label{fb_full2}
\eqnr
These equations are formidable to solve. We look for solutions
with azimuthal symmetry and also assume $v_{\theta}=0$. In the main 
text we had obtained solutions for $v_r(r)$ and $\tilde Q(r)$ by numerically solving Eq.3,4 (one can show $q=0$, see below). We showed that, in steady-state 
($\dfrac{D}{Dt}Q_{\alpha\beta}=0$), with the boundary condition that the nematics are perfectly aligned with the inner perimeter ($r=R_0$), the solutions for $\tilde Q$ are very weakly dependent on $\beta_1$, the
 flow coupling parameter. However, $v_r(r)$ strongly depends on $\beta_1$,
 via viscosity renormalisation. Therefore we simplify Eq.4 by setting
 $\beta_1=0$ and get $H_{\alpha\beta}=0$.
% $ from the $\beta_1=0$ in the  $\tilde Q$ equation limit approximation ($\beta_1\simeq0$) in Eq.~\ref{ce_2} gives $H_{\alpha\beta}=0$, 
Further using Landau-De Gennes energy functional (mentioned 
in the maintext), for  $H_{\alpha\beta}$ we get,
\beq
\label{Q_eqn}
\nabla^2 \tilde{Q}=\frac{\tilde{Q}}{L_c^2},~~\nabla^2 q=\frac{q}{L_c^2},
\eqn
%which is Eq.~5 in the maintext. 
Here $L_c = \sqrt{L/\chi}$ is the nematic correlation length scale. Converting Eq.~\ref{Q_eqn} into polar form we get (derivation given in section-4)
\beqr
\label{Qp_1}
&&\frac{1}{r}\frac{\partial}{\partial r}\left(r\frac{\partial \tilde{Q}}{\partial r}\right) -\left(\frac{1}{L_c^2}+\frac{4}{r^2}\right) \tilde{Q} = 0 \label{Q bessel}\\
&&\frac{1}{r}\frac{\partial}{\partial r}\left(r\frac{\partial q}{\partial r}\right) -\left(\frac{1}{L_c^2}+\frac{4}{r^2}\right)q = 0 \label{q evol 1}
\label{Qp_2}
\eqnr
%Here we are looking for a rotationally symmetric solution, i.e.,~there is no $\theta$-dependance. As a result 
The force-balance equations, Eqs.~(\ref{fb_full1}--\ref{fb_full2}), further simplifies to:
\beqr
\label{fb_rs1}
&&4\eta\frac{\partial}{\partial r}\left(\frac{\partial}{\partial r} + \frac{1}{r}\right)v_r + \zeta\Delta\mu\left(\frac{\partial}{\partial r}+\frac{2}{r}\right)\tilde{Q} = \alpha v_r \\
&&\frac{\partial q}{\partial r} + \frac{2q}{r} \label{q evol 2} = 0
\label{fb_rs2}
\eqnr
Eqs.~(\ref{Qp_1}--\ref{Qp_2}) and Eqs.~(\ref{fb_rs1}--\ref{fb_rs2}) are the main equations used in our study for $\beta_1=0$. We now look at several limiting cases using these set of equations.

\subsection{2.1. Nematic order with $\beta_1=0$.}
%in the weak flow coupling limit}
\label{sec_op_nfc}
Substituting Eq.~\ref{fb_rs2} in Eq.~\ref{Qp_2}, we get $q=0$. Furthermore we notice that Eq.~\ref{Qp_1} is a modified Bessel equation of order 2 \cite{simmons}. This has solution of the form
\beq
\tilde{Q}(r) = c_1 K_2(r/L_c) + c_2 I_2(r/L_c),
\eqn
where $I_2$ and $K_2$ are the modified Bessel functions of the second kind \cite{simmons}. Since $I_2$ blows up as $r\to \infty$, we set $c_2=0$. Using the other boundary condition i.e., ~$\tilde{Q}(r= R_0) = - \dfrac{1}{2}$ (see Sec.~\ref{op}), we get $c_1=-\dfrac{1}{2K_2(R_0/L_c)}$. The zeroth-order solution for the nematic order parameter is then given by 
\beq
\label{Q0}
\tilde {Q}^0(r) = -\frac{K_2(r/L_c)}{2K_2(R_0/L_c)},
\eqn
which is Eq.~5 in the main-text. \\

Instead of having the outer boundary at infinity, if it is located at $r=r_0>R_0$ (which is the realistic case since at the division plane
the cell has a finite cross-section), then using $\tilde Q=0$ at $r=r_0$ (keeping the other boundary condition
at $R_0$ unchanged), we get
\beq
\tilde Q^0(r) = 
-\frac{ I_ 2 (r) K_ 2(r_0) - K_ 2 (r) I_ 2 (r_0) }
{2[ I_ 2 (R_0) K_ 2 (r_0) -  K_ 2 (R_0) I_ 2 (r_0)]},
\eqn

\subsection{2.2. Flow field in the absence of substrate friction}
\label{sec_no_fric}
In the absence of substrate friction (i.e.,~$\alpha=0$), Eq.~\ref{fb_rs1} becomes
\beq
\label{fb_rs0}
\frac{\partial}{\partial r}\left(\frac{\partial}{\partial r} + \frac{1}{r}\right)v_r + \frac{\zeta\Delta\mu}{4\eta}\left(\frac{\partial}{\partial r}+\frac{2}{r}\right)\tilde{Q} = 0.
\eqn
Substituting Eq.~\ref{Q0} in Eq.~\ref{fb_rs0}, we get a second order differential equation in $v_r$ with a source term, namely,
\beq
\label{v_nh}
\frac{\partial}{\partial r}\left(\frac{\partial}{\partial r}+\frac{1}{r}\right)v_r=-\frac{\zeta\Delta\mu}{4\eta}\frac{K_1(r/L_c)}{2K_2(R_0/L_c)}.
\eqn
The homogeneous solution of Eq.~\ref{v_nh} is of the form $(A r+B/r)$. To obtain the particular solution, we use an ansatz $v_p(r)=\Omega K_1(r/L_c)$. This is inspired by the form of the source term in Eq.~\ref{v_nh}. Substituting this particular solution in Eq.~\ref{v_nh} gives us $\Omega = -\dfrac{\zeta\Delta\mu}{4\eta}\dfrac{L_c}{2K_2(R_0/L_c)}$. The general solution for $v_r$ is then given by 
\beq
\label{vr_gen}
v_r(r) = A r+\frac{B}{r} -\frac{\zeta\Delta\mu}{4\eta}\frac{L_c}{2K_2(R_0/L_c)} K_1(r/L_c).
\eqn 
We put $A=0$, since $v_r\to 0$ as $r\to\infty$. The constant $B$ is calculated using the free boundary condition at $r=R_0$, i.e. $\sigma_{rr}(R_0)=0$. Using this condition in Eq.~\ref{stress_p1}, we get a boundary condition for $v_r'$ at $R_0$, which is given by
\beq
\label{vrp_bc}
 2\eta v_r'(R_0)+\dfrac{\zeta\Delta\mu}{6}-\dfrac{\zeta\Delta\mu}{2}=0. 
\eqn
 Using Eq.~\ref{vr_gen} and Eq.~\ref{vrp_bc} we get $B=-\dfrac{\zeta\Delta\mu R_0^2}{6\eta}\left[1+\dfrac{K_1'(R_0/L_c)}{K_2(R_0/L_c)}\right]$, which leads to the final expression for the zeroth-order velocity field $v_r^0(r)$ (Eq.~6 in the maintext), namely,
\beq
v_r^0(r) = -\frac{\zeta\Delta\mu}{\eta}\Bigg[\left(1+\frac{3K_1'(R_0/L_c)}{4K_2(R_0/L_c)}\right)\frac{R_0^2}{6r} + \frac{L_c}{8}\frac{K_1(r/L_c)}{K_2(R_0/L_c)}\Bigg] 
\label{zeroth_v}
\eqn

The corresponding velocity equation, for the case of outer boundary located at finite $r=r_0$, and $v_r(r_0)=0$ (keeping the other stress boundary condition at $r=R_0$ unchanged) is solved numerically using Mathematica.

\subsection{2.3. Flow-field in the presence of substrate friction}
\label{sec_friction}
Here we consider the rotationally symmetric case but in the presence of substrate friction ($\alpha \neq0$). The force-balance equation given by Eq.~\ref{fb_rs1} can be re-written as
\beq
\label{fb_wf}
\underbrace{\left[\partial_r\left(\partial_r+\frac{1}{r}\right) - \tilde{\alpha}\right]}_{\mathcal{L}}v_r = f(r).
\eqn
Here $ \tilde{\alpha } = \dfrac{\alpha }{4\eta}$ and $f(r)=-\dfrac{\zeta\Delta\mu}{4\eta}\left(\partial_r+\dfrac{2}{r}\right)\tilde{Q}_{0}  = -\dfrac{\zeta\Delta\mu}{4\eta} \dfrac{K_1(r/L_c)}{2K_2(R_0/L_c)}$. Here $\tilde{Q}_{0}(r)$ is the zeroth-order solution obtained for $\beta_1=0$ (see Eq.~\ref{Q0}).  Eq.~\ref{fb_wf} is a linear non-homogenous differential equation and can be solved using the method of Green's function with boundary conditions $v_r(r\to\infty) = 0$, and $\sigma_{rr}(R_0) = 0$.

\medskip 

The Green's function $G(r,r')$ for the differential operator $\mathcal{L}$ appearing in Eq.~\ref{fb_wf} satisfies the differential equation:
\beq
\label{green_fn1}
\mathcal{L}G(r,r')=\left[\frac{\partial^2}{\partial r^2}+\frac{1}{r}\frac{\partial}{\partial r}-\frac{1}{r^2}(\tilde{\alpha}r^2+1)\right]G(r,r') = \delta(r-r').
\eqn
Eq.~\ref{green_fn1} is a modified Bessel equation of first order. This general solution is written in terms of the modified Bessel functions of first and second kind.
\beqr
\label{gfn_sol1}
r < r'  &&: \quad G(r,r') = G_<(r,r') =A_1I_1(\sqrt{\tilde{\alpha}}r)+A_2K_1(\sqrt{\tilde{\alpha }}r), \\
r > r' &&: \quad G(r,r') = G_>(r,r') =B_1 I_1(\sqrt{\tilde{\alpha }}r)+ B_2K_1(\sqrt{\tilde{\alpha}}r).
\label{gfn_sol2}
\eqnr

The full solution to Eq.~\ref{fb_wf} can be written as,
\beq
\label{gfn_full1}
v_r(r) = v_h(r) + \int_{R_0}^\infty  G(r,r')f(r')dr'.
\eqn
Here $v_h(r)$ represents the homogeneous solution obtained by solving $ \mathcal{L}v_r = 0$. It is easy to check that it is given by
\beq
\label{vh_fr}
v_h(r) =-\frac{\zeta \Delta\mu}{3\eta\sqrt{\tilde{\alpha }}}\frac{K_1(\sqrt{\tilde{\alpha }}r)}{K_0(R_0\sqrt{\tilde{\alpha }})+K_2(R_0\sqrt{\tilde{\alpha }})} 
\eqn

The unknown constants in Eqs.~(\ref{gfn_sol1}-\ref{gfn_sol2}) are determined from the boundary conditions. Here $G(r,r')$ is defined in the range $[R_0,\infty)$. By using the known boundary conditions for $v_r$ and substituting the expression of Eq.~\ref{vh_fr} in Eq.~\ref{gfn_full1}, we notice that $G(r,r')$ should satisfy the following boundary conditions at $r=R_0$ and $ r\to\infty$ respectively
\beqr
\label{gfn_bc1}
&& \partial_r G(r,r')\Big|_{r=R_0} = 0, \label{bc1 green's fn} \\
&& G(r\to\infty,r') = 0. \label{bc2 green's fn}
\label{gfn_bc2}
\eqnr

Eqs.~(\ref{bc2 green's fn},\ref{bc1 green's fn}) give $B_1 = 0$ and $A_1 = A_2\frac{K_0(\sqrt{\tilde{\alpha }}R_0)+ K_2(\sqrt{\tilde{\alpha }}R_0)}{I_0(\sqrt{\tilde{\alpha }}R_0)+ I_2(\sqrt{\tilde{\alpha }}R_0)}$, respectively. The remaining two boundary conditions are obtained by integrating Eq.~\ref{green_fn1} twice over an infinitesimally small interval around $r'$. The gives the continuity conditions of $G(r,r')$ and its derivative at $r=r'$, namely,
\beqr
\label{gfn_bc3}
&& \frac{\partial G_>}{\partial r}\Big|_{r\to r'_+} - \frac{\partial G_<}{\partial r}\Big|_{r\to r'_-} = 1, \\
&& G_<(r,r')\Big|_{r \to r'_-}= G_>(r,r')\Big|_{r\to r'_+}.
\label{gfn_bc4}
\eqnr
\newline

Eqs.(~\ref{gfn_bc3},\ref{gfn_bc4}) give $B_2 = \frac{A_2 \left(K_0\left(\sqrt{\tilde{\alpha }} R_0\right)+K_2\left(\sqrt{\tilde{\alpha } } R_0\right)\right)
   I_1\left(r' \sqrt{\tilde{\alpha }}\right)}{\left(I_0\left(\sqrt{\tilde{\alpha }}
   R_0\right)+I_2\left(\sqrt{\tilde{\alpha }} R_0\right)\right) K_1\left(r' \sqrt{\tilde{\alpha }}\right)}+A_2$, where 
   \\ $A_2 = -\frac{2. \left(I_0\left(\sqrt{\tilde{\alpha } } R_0\right)+I_2\left(\sqrt{\tilde{\alpha } }
   R_0\right)\right)}{\sqrt{\tilde{\alpha } } r' \left(K_0\left(\sqrt{\tilde{\alpha } }
   R_0\right)+K_2\left(\sqrt{\tilde{\alpha } } R_0\right)\right) \left(K_0\left(r' \sqrt{\tilde{\alpha }
   }\right)+K_2\left(r' \sqrt{\tilde{\alpha } }\right)\right) \left(\frac{I_1\left(r' \sqrt{\tilde{\alpha }
   }\right)}{K_1\left(r' \sqrt{\tilde{\alpha } }\right)}+\frac{\left(I_0\left(r' \sqrt{\tilde{\alpha }
   }\right)+I_2\left(r' \sqrt{\tilde{\alpha } }\right)\right)}{K_0\left(r' \sqrt{\tilde{\alpha }
   }\right)+K_2\left(r' \sqrt{\tilde{\alpha } }\right)}\right)}$  \newline

The full solution is now written as:
\beq
\label{gfn_full2}
v_r(r) =  -\frac{\zeta \Delta\mu}{3\eta\sqrt{\tilde{\alpha }}}\frac{K_1(\sqrt{\tilde{\alpha }}r)}{K_0(R_0\sqrt{\tilde{\alpha }})+K_2(R_0\sqrt{\tilde{\alpha }})} + \int_{R_0}^r G_>(r,r')f(r')dr' + \int_{r}^\infty G_<(r,r')f(r')dr'.
\eqn
%One can check that Eq.~\ref{gfn_full2} satisfies Eq.~\ref{fb_wf} subject to the boundary conditions at $ r=R_0 $ and $ r\to\infty $. 
%\red{Comment: Please put a short statement on how the last two terms in Eq.~\ref{gfn_full2} is obtained numerically using Mathematica.} \\

The two integrals in Eq.~\ref{gfn_full2} are evaluated 
numerically, for different values of $r$, using "NIntegrate" 
in Mathematica. The constants $R_0, \alpha$ are kept fixed.
In this manner we obtain a table for $v_r$ (in units of 
$\frac{\zeta \Delta\mu}{\eta}$) versus $r$.
%   once we obtain the corresponding functions and then plotted the values obtained numerically to get the velocity.

\section{3. Correction to the nematic Field due to non-zero $\beta_1$}
\label{sec_flowcoupling}
%The dynamical equations for nonzero $\beta_1$ are already given in
%the maintext as Eq.16 and 17, and their numerical solutions (with
%$\alpha=0$) are presented in Fig.4.  
We have already shown numerically (main-text) that the correction to 
$\tilde Q(r)$ is small even for finite $\beta_1$. This correction
can be estimated for small $\beta_1$ 
%But for weak flow copuling, i.e., ~$\beta_1<<1$ we can compute $\tilde Q(r)$ 
perturbatively, up to different orders in $\beta_1$, using Greens function approach. For that we use the zeroth order solution  $v_r^0$
(Eq.\ref{zeroth_v} with bare viscosity $\eta$, corresponding to $\beta_1=0$), on the right hand side of Eq.3, in the main-text, which now reads,
%We now consider the limiting case where the flow is still rotationally symmetric but is weakly coupled to the nematic field, i.e., ~$\beta_1$ is small but $\neq0$. Neglecting the substrate friction ($\alpha=0$), the steady-state $\tilde{Q}$ equation is now modified to
\beq
\partial_r\left(r\partial_r\right)\tilde{Q}-\left(\frac{r}{L_c^2}+\frac{4}{r}\right) \tilde{Q} = -\frac{1}{2L} \beta _1 \beta _2 r\left(\partial_r  v_r^0 - \frac{ v_r^0}{r}\right).\label{Q-smallbeta}
\eqn
%Substituting the zeroth-order solution for the flow-field $v_r=$
Substituting for $v_r^0$ %from Eq.~\ref{zeroth_v},% 
we get,
\beq
\partial_r\left(r\partial_r\right)\tilde{Q}-\frac{r}{L_c^2}\tilde{Q} - \frac{4}{r}\tilde{Q} =-\frac{\beta _1 \beta _2 r}{2L}\left(\frac{R_0^2 \left(5-\frac{3 K_0(R_0/L_c)}{K_2(R_0/L_c)}\right)}{24 r^2}+\frac{\frac{L_C}{r} K_1(r/L_c)+ \frac{1}{2} (K_0(r/L_c)+K_2(r/L_c))}{8 K_2(R_0/L_c)}\right)  
\eqn
which can be re-written as follows
\beq
\label{Q_fc}
\underbrace{\left[\partial_r\left(r\partial_r\right)-\frac{r}{L_c^2} - \frac{4}{r}\right]}_{\mathcal{L}}\tilde{Q} = u(r),
\eqn
where $u(r)$ is the right hand side of Eq.\ref{Q-smallbeta}.
%with $u(r)=-\frac{\beta _1 \beta _2 r}{2}\left(\frac{R_0^2 \left(5-\frac{3 K_1(R_0/L_c)}{K_2(R_0/L_c)}\right)}{24 r^2}+\frac{\frac{L_C}{r} K_1(r/L_c)+0.5 (K_0(r/L_c)+K_2(r/L_c))}{8 K_2(R_0/L_c)}\right)$. 
The boundary conditions on $\tilde{Q}$ are: $\tilde{Q}(r\to\infty) = 0$, and $\tilde{Q}(r\to R_0) = -0.5$. The Green's function $G(r,r')$ for operator $\mathcal{L}$ in Eq.~\ref{Q_fc} satisfies the modified Bessel equation of order 2.
\beq
\label{Qgfn}
\mathcal{L}G(r,r') = \left[\partial_r\left(r\partial_r\right)-\frac{r}{L_c^2} - \frac{4}{r}\right]G(r,r') = \delta(r-r'),
\eqn
having a solution of the form
\beqr
r<r':&& \quad G(r,r') = G_<(r,r') =A_1I_2(r)+A_2K_2(r), \\
r>r':&& \quad G(r,r') = G_>(r,r') =B_1I_2(r)+B_2K_2(r).
\eqnr
The four unknown constants can be determined from the boundary conditions on $G(r,r')$ which as before are given by,
\beqr
&& G(r\to R_0,r') = 0 \label{bc1 q beta}, \\
&& G(r\to\infty,r') = 0 \label{bc2 q beta}, \\
&& \frac{\partial G_>}{\partial r}\Big|_{r\to r'_+} - \frac{\partial G_<}{\partial r}\Big|_{r\to r'_-} = 1, \label{bc3 q beta} \\
&& G_<(r,r')\Big|_{r \to r'_-}= G_>(r,r')\Big|_{r\to r'_+} \label{bc4 q beta}.
\eqnr

Eq.~(\ref{bc1 q beta}--\ref{bc4 q beta}) gives $B_1 = 0$, $A_2 =-\frac{A_1I_2(R_0)}{K_2(R_0)}$, $B_2 =A_1\left(\frac{I_2(r')}{K_2(r')}-\frac{I_2(R_0)}{K_2(R_0)}\right)$, and $A_1=\frac{K_2(r')}{r'(I_2(r')K_2'(r') - I_2'(r')K_2(r'))}$. 
%\red{Comment: %Please check, which one is $A_1$ and which one is $A_2$? \blue{done sir}} \\

The full solution to Eq.~\ref{Q_fc} is therefore
\beq
\label{Qgfn_sol}
\tilde{Q}(r) = \tilde{Q}_h(r) + \int_{R_0}^\infty  G(r,r')u(r')dr',
\eqn
where $\tilde{Q}_h(r)=-\dfrac{K_2(r/L_c)}{2K_2(R_0/L_c)}$ is homogeneous solution obtained by solving $\mathcal{L}\tilde{Q}=0$. The full solution can then be rewritten as:
\beq
\tilde{Q} = -\dfrac{K_2(r/L_c)}{2K_2(R_0/L_c)} + \int_{R_0}^r G_>(r,r')u(r')dr' + \int_{r}^\infty G_<(r,r')u(r')dr'.
\eqn

The corresponding velocity equation (Eq.\ref{fb_rs1}) is easy to handle in the small $\beta_1$ limit. Substituting $\tilde Q^0$ in place of 
$\tilde Q$ gives us
\beq
\label{fb_fc1}
4\eta\partial_r\left(\partial_r+\frac{1}{r}\right)v_r=
-\frac{\zeta\Delta\mu}{1+\frac{\beta_1^2\beta_2}{2\eta}}\left(\partial_r+\frac{2}{r}\right)\tilde{Q}_{0}  
\eqn
Note that this equation has the same structure as Eq.\ref{fb_rs0}, with
the viscosity $\eta$ enhanced by a factor of 
$(1+\frac{\beta_1^2\beta_2}{2\eta})$. Therefore the solution is
same as Eq.\ref{zeroth_v}, where $\eta$ has to be replaced by $\eta(1+\frac{\beta_1^2\beta_2}{2\eta})$.

\section{4. Derivation of dynamical equations for $\tilde Q,q$ (Eq.\ref{Qp_1},\ref{Qp_2} from Eq.\ref{Q_eqn})}
 The polar form of $Q'_{\alpha\beta}$ matrix,
$\begin{bmatrix}
Q'_{rr}&Q'_{r\theta}\\
Q'_{\theta r}&Q'_{\theta\theta}
\end{bmatrix}$,
is given by 
\begin{align}
\begin{bmatrix}
\cos{\theta}&\sin{\theta}\\
-\sin{\theta}&\cos{\theta}
\end{bmatrix}
\begin{bmatrix}
\tilde Q&q\\
q&-\tilde Q
\end{bmatrix}
%\begin{bmatrix} 
%-\frac{\cos{2\theta}}{2}&-\frac{\sin{2\theta}}{2}\\ -\frac{\sin{2\theta}}{2}&\frac{\cos{2\theta}}{2}
%\end{bmatrix}
\begin{bmatrix}
\cos{\theta}&-\sin{\theta}\\
\sin{\theta}&\cos{\theta}
\end{bmatrix}
=
\begin{bmatrix}
\tilde Q\cos{2\theta} + q\sin{2\theta}&q\cos{2\theta} - 
\tilde Q\sin{2\theta}\\
q\cos{2\theta} - \tilde Q\sin{2\theta}&-(\tilde Q\cos{2\theta} + q\sin{2\theta})
\end{bmatrix}
\equiv
\begin{bmatrix}
Q_1&q_1\\
q_1&-Q_1
\end{bmatrix} 
\end{align}
\newline
where $\begin{bmatrix}
\tilde Q&q\\
q&-\tilde Q
\end{bmatrix}$ is its corresponding cartesian form (in $x-y)$.
Noting that the basic structure of the matrix is retained, we 
express the new matrix elements in terms of the old ones as,  
$Q_1 = \tilde Q\cos{2\theta} + q\sin{2\theta}$ and,
$q_1 = - \tilde Q\sin{2\theta} + q\cos{2\theta}$.
Inverting these we get, 
$\tilde Q = Q_1\cos{2\theta} - q_1\sin{2\theta}$ and,
$q = Q_1\sin{2\theta} + q_1\cos{2\theta}$. Note that $Q_1,q_1$ 
are functions of $r$ via $\tilde Q,q$. 
Now substituting the polar form for the laplacian operator
and $\tilde Q$ into Eq.\ref{Q_eqn}, $\nabla^2 \tilde Q=
\frac{\tilde Q}{L_c^2}$ (whose polar form we are interested in), we get
\begin{eqnarray}
%(\nabla_r^2+\frac{\nabla_\theta^2}{r^2})Q \equiv 
\Big[\frac{1}{r}\frac{\partial}{\partial r}\left(r\frac{\partial}{\partial r}\right) + \frac{1}{r^2}\frac{\partial^2}{\partial \theta^2}\Big](Q_1\cos{2\theta} - q_1\sin{2\theta}) &=&
\frac{Q_1\cos{2\theta} - q_1\sin{2\theta}}{L_c^2}\nonumber\\
%------------
\frac{1}{r}\frac{\partial}{\partial r}\left(r\frac{\partial Q_1}{\partial r}\right)\cos{2\theta} -  \frac{1}{r}\frac{\partial}{\partial r}\left(r\frac{\partial q_1}{\partial r}\right)\sin{2\theta}
 -\frac{Q_1}{r^2}4\cos{2\theta} + \frac{q_1}{r^2}4\sin{2\theta} &=& \frac{Q_1\cos{2\theta} - q_1\sin{2\theta}}{L_c^2}
%------------
\end{eqnarray}

%Substitute Q :\\
%$(\nabla_r^2+\frac{\nabla_\theta^2}{r^2})(Q_1\cos{2\theta} - q_1\sin{2\theta}) = \frac{Q_1\cos{2\theta} - q_1\sin{2\theta}}{L_c^2}$ \\
%\newline
Further, separating the coefficients of $\cos 2\theta$ and 
$\sin 2\theta$ from the two sides, yields 
\begin{equation}
\Large[\frac{1}{r}\frac{\partial}{\partial r}\left(r\frac{\partial Q_1}{\partial r}\right) - 4\frac{Q_1}{r^2}-\frac{Q_1}{L_c^2}\large]\cos{2\theta} -
[\frac{1}{r}\frac{\partial}{\partial r}\left(r\frac{\partial q_1}{\partial r}\right) - 4\frac{q_1}{r^2}- \frac{q_1}{L_c^2}]\sin{2\theta}
= 0
\end{equation}
%or,\newline
%$\cos{2\theta}(\nabla_r^2 Q_1) - \sin{2\theta}(\nabla_r^2 q_1) -\frac{Q_1}{r^2}4\cos{2\theta} + \frac{q_1}{r^2}4\sin{2\theta} = \frac{Q_1\cos{2\theta} - q_1\sin{2\theta}}{L_c^2}$\\
%\newline
%or,\newline
%$(\nabla_r^2 Q_1  - 4\frac{Q_1}{r^2})\cos{2\theta} - (\nabla_r^2 q_1  - 4\frac{q_1}{r^2})\sin{2\theta} = \frac{Q_1\cos{2\theta} - q_1\sin{2\theta}}{L_c^2}$ \\
%\newline
Since $\cos{2\theta} $ and $  \sin{2\theta}$ are independent fucntions their coefficients, separately, must be zero. Thus we 
obtain the order parameter equations Eq.\ref{Qp_1} and Eq.\ref{Qp_2}, where we retained the same notations as in the cartesian frame, by changing $(Q_1,q_1)\rightarrow (\tilde Q,q)$. 
%\newline
%$\nabla_r^2 Q_1  - 4\frac{Q_1}{r^2} = \frac{Q_1}{L_c^2}$ \\
%$\nabla_r^2 q_1  - 4\frac{q_1}{r^2} = \frac{q_1}{L_c^2}$ \\
%\newline

\section{5. Computation of closure rate $R_0(t)$ vs $t$.}

This requires the velocity at the ring $v_r(R_0)$ for different
values of $R_0$. When the outer radius is located at infinity
we have a closed form expression for $v_r(R_0)$ which can be 
integrated to obtain the closure rate. But for the case of
finite outer boundary (at $r_0$) we could solve the velocity equation (Eq.18) only numerically (using "NDsolve" in Mathematica)  after incorporating 
the analytic expression for $\tilde Q^0(r)$ (Eq.22). Therefore we first prepared a table for  $v_r(R_0)$ versus $R_0$ by solving the velocity equation for different values of $R_0$. We then obtained an interpolation function for this table. This function is further
integrated numerically  to obtain $R_0(t)$ versus $t$.

%get $v_r(R_0)$ to solve the
%Once we get $Q_0(r) = \frac{K_ 2 (r) I_ 2 (r_0) - I_ 2 (r) K_ 2(r_0)}{2 I_ 2 (R_0) K_ 2 (r_0) - 2 K_ 2 (R_0) I_ 2 (r_0)},$ using eqn $\frac{1}{r}\frac{\partial}{\partial r}\left(r\frac{\partial \tilde{Q}}{\partial r}\right) -\left(\frac{1}{L_c^2}+\frac{4}{r^2}\right) \tilde{Q} = 0$ we obtain velocity numerically in mathematica using NDSolve using eqn $4\eta\frac{\partial}{\partial r}\left(\frac{\partial}{\partial r} + \frac{1}{r}\right)v_r + \zeta\Delta\mu\left(\frac{\partial}{\partial r}+\frac{2}{r}\right)\tilde{Q} = 0$ with stress free and velocity at $r = r_0$ as boundary conditions.\\
%Now a table comprising values for $v(R_0)$ $ vs $ $ R_0$ for different values of $R_0$ is made.\\ Once we get a closed form for $v(R_0)$ using interpolation we again use NDSolve to numerically obtain $R_0(t) $ $ vs $ $ t$.  \\
{\bf  All numerical calculations have been performed using Mathematica version 11.3}.

%opening
\title{Supplementary Information \underline{Part-II} : Dynamics and stability of the cytokinetic ring }
%\title{Linear stability analysis for asymmetric contraction of the cytokinetic ring}
%\input author_list.tex
\author{Mainak Chatterjee, Arkya Chatterjee, Amitabha Nandi, and Anirban Sain\\\textit{Department of Physics, Indian Institute of Technology-Bombay, Powai, Mumbai 400076, India}}  
\date{\today}
%\begin{abstract}
%	The contraction of the cytokinetic ring is a crucial part of the cell-division cycle. In a the main text and Supplementary Information part-1 (SI-1) 
	%previous work \cite{CCNS2007.13441}, 
%	using a continuum gel theory framework in a weak flow coupling regime, we obtained exact analytical solutions for the quasi-static dynamics of the radially symmetric ring contraction. Here, we study the stability of this process under angular perturbations. Linear stability analysis of the angular modes shows that, in the experimentally relevant regime of the parameter space, the lowest order breathing mode and most of the higher-order modes are stable, leaving a window of unstable modes in between. Furthermore as the ring shrinks in radius, this window gradually disappears, turning all modes stable. This is consistent with the experimental observation \cite{MLDODevCell07} that the ring, most often, is not circular at the beginning, but becomes increasingly circular as it contracts.
%\end{abstract}

%\pacs{87.16.Ka, 87.16.ad, 87.16.dj, 87.17.Ee}
\maketitle

%One of the problems that we would like to understand is the constriction of the contractile ring of actomyosin during cytokinesis. The experimental paper \cite{MLDODevCell07} demonstrates the robustness of this ring closure phenomenon, wherein the time of closure of the cortical ring formed at the time of cytokinesis remains unchanged even after laser ablation of part of the ring. We suspect that there might be a physical explanation behind this robustness that allows timely completion of cytokinesis.
%
%In \cite{ring}, the authors discuss an active gel model of the cytoskeletal flows leading to ring formation leading up to cytokinesis. An effective  2-D theory is developed in a rotationally symmetric setting to obtain an estimate for the ring closure time. As stated above, the experimental paper \cite{MLDODevCell07} seems to suggest that this ring closure phenomenon is quite robust. In the following section(s), we adopt a slightly different approach to first get to a zeroth order (rotationally symmetric) solution and subsequently show that we can do a linear stability analysis to show that this solution is stable against fluctuations.

\begin{figure}
	\centering
	\includegraphics[width=\linewidth]{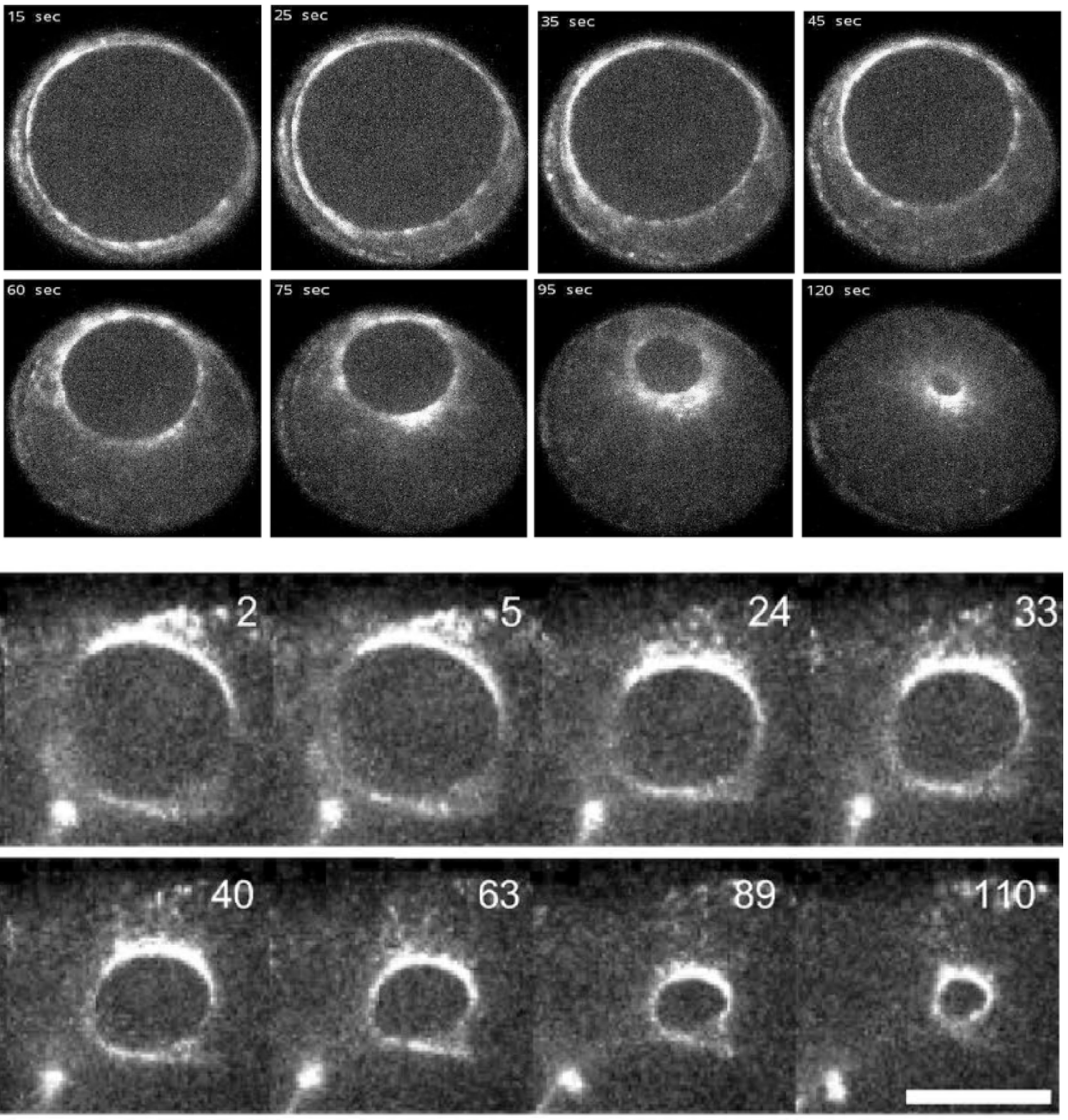}
	\caption{Experimental images of ring closure, in c Elegans cell division, adapted from different sets of experiments by Menon et al. \cite{vidya} (the top two panels) and Silva et al.\cite{SOPM..JCellBiol16} (the 3rd and 4th panels). The scale bar is 5 $\mu$m (for the lower panel), and time is in seconds.} %Time zero denotes the frame before the laser cut is made.}
	\label{fig:laser_expt}
\end{figure}

%\section{Stability analysis of ring deformation}
In the main text and Supplementary material part-1 (SI-I) %\cite{CCNS2007.13441}, 
we studied the contraction of the cytokinetic ring with radial symmetry as a simplifying assumption. Here, we extend the analysis to a deformed hole in order to investigate the stability of the symmetric solution derived previously. The motivation for this study 
arises from experimental observation about the typical shape of the ring during 
cytokinesis (see Fig-\ref{fig:laser_expt} and Ref\cite{MLDODevCell07, SOPM..JCellBiol16,vidya}). 
As shown in Fig.\ref{fig:laser_expt}, the ring contraction does not occur in a radially symmetric fashion. At the outset, the ring appears to be far from a perfect circle, but as it contracts, it becomes increasingly circular. We would like to address this observation by studying the stability of the different angular modes of the contraction dynamics. In order to make this exercise analytically tractable (in the sense of linear stability analysis), we assume small deviation from the circular shape. This is a strategy analogous to that adopted in \cite{PABG...NatPhys19} and \cite{alert_thesis} (chapter-7). By using linear stability analysis for the quasi-static ring contraction dynamics, we are able to calculate the growth/decay rate of the angular modes using perturbation theory. We find that the lowest-order breathing mode and most of the higher-order modes are stable for experimentally relevant regime of parameter values. Furthermore, for large values of ring radius, there is a window of unstable modes in between; however this window shrinks as the ring contracts and eventually all modes become stable.

We study deformations of the circular ring by decomposing into Fourier modes as modes as follows:
\begin{equation}
\delta R(\theta,t)=\sum_{n=0}^{\infty} \delta R_n(t) e^{in\theta}
\end{equation}
This perturbative deformation to the ring implies that the circular ring defined by the polar equation $r(\theta)=R_0$ is now changed to $r(\theta)=R_0+\delta R(\theta)$ (the time dependence is suppressed for simplicity of notation). 

\begin{figure}
	\centering
	\includegraphics[width=0.7\linewidth]{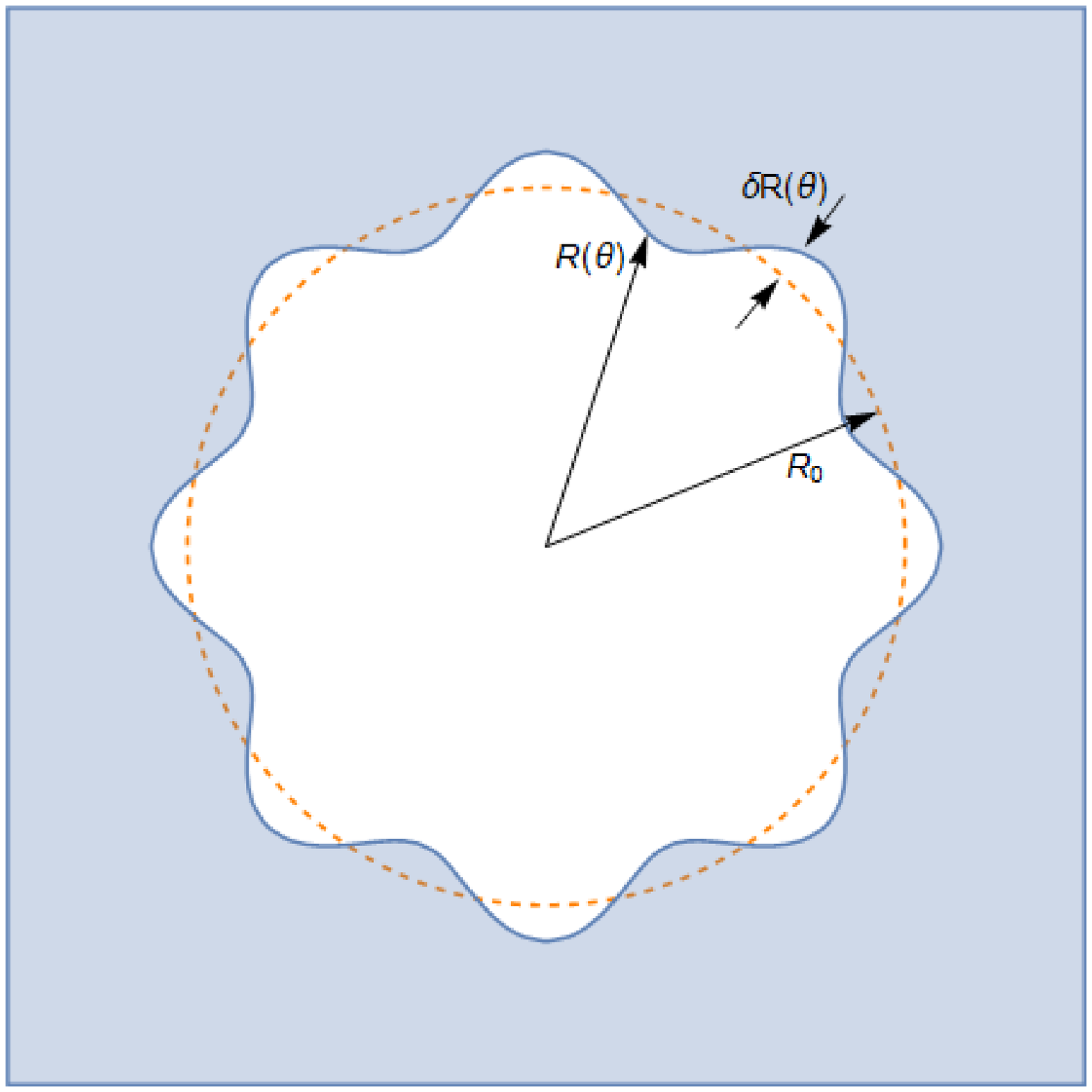}
	\caption{Deformed hole}
	\label{fig:deformed_ring}
\end{figure}
Due to this deformation, velocity and order parameter fields change perturbatively:
\begin{align}
	\tilde{Q}(r,\theta,t) &=\tilde{Q}_0(r)+\delta\tilde{Q}(r,\theta,t)\label{Qexpnd}\\
	q(r,\theta,t) &=\delta q(r,\theta,t)\label{qexpnd}\\
	v_r(r,\theta,t) &=v_r^0(r)+\delta v_r(r,\theta,t)\\
	v_\theta(r,\theta,t) &=\delta v_\theta(r,\theta,t)
\end{align}

The perturbation fields $\delta \tilde{Q}, \delta q, \delta v_{r,\theta}$ can be expanded similar to $\delta R(\theta,t)$ as
\begin{align}
	\delta\tilde{Q}(r,\theta,t) &= \sum_{n=0}^{\infty} \delta \tilde{Q}_n(r,t) e^{in\theta}\label{Qmodes}\\
	\delta q(r,\theta,t) &= \sum_{n=0}^{\infty} \delta q_n(r,t) e^{in\theta}\label{qmodes}\\
	\delta v_r(r,\theta,t) &= \sum_{n=0}^{\infty} \delta v_{r,n}(r,t) e^{in\theta}\\
	\delta v_\theta(r,\theta,t) &= \sum_{n=0}^{\infty} \delta v_{\theta,n}(r,t) e^{in\theta}
\end{align}
We substitute these expansions in the dynamical equations and use the exact forms of the zeroth order solutions to solve for the Fourier amplitudes of all the four fields.

%\begin{figure*}
%	\centering
%	\includegraphics[width=\textwidth]{fourier_modes.png}
%	\caption{Various Fourier modes with undeformed ring in dotted lines}
%	\label{fig:fourier_modes}
%\end{figure*}

\textit{Dynamical equations for $\delta\tilde{Q}$ and $\delta q$} ---
The Fourier amplitudes $\delta\tilde{Q}_n$ and $\delta q_n$ of the order parameter fields $\delta\tilde{Q}$ and $\delta q$ satisfy the following equations:
\begin{align}
	\left(\partial_r^2+\frac{1}{r}\partial_r-\frac{1}{L_c^2}-\frac{n^2+4}{r^2}\right)\delta \tilde{Q}_n &= \frac{4in}{r^2}\delta q_n \label{delQneq}\\
	\left(\partial_r^2+\frac{1}{r}\partial_r-\frac{1}{L_c^2}-\frac{n^2+4}{r^2}\right)\delta q_n &= -\frac{4in}{r^2}\delta \tilde{Q}_n \label{delqneq} 
\end{align}
We can diagonalise these equations by a change of variables,
\begin{align}
	\delta Q_+ &=\delta \tilde{Q}_n+i\delta q_n\\
	\delta Q_- &=i\delta \tilde{Q}_n+\delta q_n
\end{align}
In terms of the variables $ \delta Q_+ $ and $ \delta Q_- $, \eqref{delQneq} and \eqref{delqneq} give
\begin{align}
	\left(\partial_r^2+\frac{1}{r}\partial_r-\frac{1}{L_c^2}-\frac{(n+2)^2}{r^2}\right)\delta Q_+ &= 0\\
	\left(\partial_r^2+\frac{1}{r}\partial_r-\frac{1}{L_c^2}-\frac{(n-2)^2}{r^2}\right)\delta Q_- &= 0
\end{align}
These are modified Bessel equations with solutions of the form
\begin{align}
	\delta Q_+ &=c_n^+ K_{n+2}(r/L_c)\\
	\delta Q_- &=c_n^- K_{n-2}(r/L_c)
\end{align}
where we have left out the Bessel-I functions since $\delta Q_\pm$ must go to 0 as $r\to\infty$. Changing back to $\delta \tilde{Q}_n$ and $\delta q_n$, we get
\begin{align}
	\delta \tilde{Q}_n &= \frac{c_n^+ K_{n+2}(r/L_c)-c_n^- iK_{n-2}(r/L_c)}{2}\\
	\delta q_n &= \frac{c_n^- K_{n-2}(r/L_c)-i c_n^+ K_{n+2}(r/L_c)}{2}\label{qbc}
\end{align}

To evaluate the constants $c_n^\pm$, we need to use the boundary conditions for the order parameter at the deformed ring $r=R(\theta)$,
\begin{equation}
\tilde{Q}(R(\theta))=-\frac{1}{2}\quad\text{and}\quad q(R(\theta))=\frac{1}{R_0}\frac{d\delta R}{d\theta}\label{Qbc}
\end{equation}
By substituting \eqref{Qexpnd},\eqref{qexpnd}, \eqref{Qmodes} and \eqref{qmodes} in \eqref{Qbc}, and expanding both sides to lowest order, we can obtain the following boundary conditions for the perturbation fields
\begin{align}
	\delta \tilde{Q}(R_0, \theta)&=-\partial_r \tilde{Q}_0(R_0)\delta R(\theta)\\
	\delta q(R_0,\theta) &= \frac{1}{R_0}\frac{d\delta R(\theta)}{d\theta}
\end{align}
which imply, for the Fourier modes,
\begin{align}
	\delta \tilde{Q}_n(R_0)&=-\partial_r \tilde{Q}_0(R_0)\delta R_n\\
	\delta q_n(R_0) &= \frac{in}{R_0}\delta R_n
\end{align}
Using these boundary conditions, we can express the exact analytical solutions for the perturbation fields $\delta\tilde{Q}_n$ and $\delta q_n$,
\begin{widetext}
\begin{align}
\delta\tilde{Q}_n(r) &= \frac{\delta R_n}{2R_0}\Bigg[\left(n-1-\frac{R_0}{2L_c}\frac{K_1(R_0/L_c)}{K_2(R_0/L_c)}\right)\frac{K_{n-2}(r/L_c)}{K_{n-2}(R_0/L_c)} - \left(n+1+\frac{R_0}{2L_c}\frac{K_1(R_0/L_c)}{K_2(R_0/L_c)}\right)\frac{K_{n+2}(r/L_c)}{K_{n+2}(R_0/L_c)}\Bigg]\label{delQn}\\
\delta q_n(r) &= \frac{i\delta R_n}{2R_0}\Bigg[\left(n-1-\frac{R_0}{2L_c}\frac{K_1(R_0/L_c)}{K_2(R_0/L_c)}\right)\frac{K_{n-2}(r/L_c)}{K_{n-2}(R_0/L_c)} + \left(n+1+\frac{R_0}{2L_c}\frac{K_1(R_0/L_c)}{K_2(R_0/L_c)}\right)\frac{K_{n+2}(r/L_c)}{K_{n+2}(R_0/L_c)}\Bigg]\label{delqn}
\end{align}
\end{widetext}

\textit{Dynamical equations for $\delta v_r$ and $\delta v_\theta$} ---
As discussed in the main text and SI-I, the velocity dynamics is expressed in terms of Navier-Stokes-like equations for the stress tensor. To perform our perturbative analysis, we expand those equations in terms of the perturbation fields $\delta v_r$ and $\delta v_\theta$ and then expand in terms of their Fourier modes. After some algebra, we can find the following dynamical equations for $\delta v_{r,n}$ and $\delta v_{\theta,n}$:
\begin{widetext}
\begin{align}
4\eta\left(\partial_r^2 +\frac{1}{r}\partial_r-\frac{4+n^2}{4r^2}\right)\delta v_{r,n}+in\eta\left(\frac{3}{r}\partial_r-\frac{5}{r^2}\right)\delta v_{\theta,n}
+\zeta\Delta\mu\left[\left(\partial_r+\frac{2}{r}\right)\delta\tilde{Q}_n + \frac{in\delta q_n}{r}\right] &=0\label{delvr}\\
\eta\left(\partial_r^2 +\frac{1}{r}\partial_r-\frac{1+4n^2}{r^2}\right)\delta v_{\theta,n}+in\eta\left(\frac{3}{r}\partial_r+\frac{5}{r^2}\right)\delta v_{r,n}
+\zeta\Delta\mu\left[\left(\partial_r+\frac{2}{r}\right)\delta q_n - \frac{in\delta\tilde{Q}_n}{r}\right] &=0\label{delvtheta}
\end{align}
\end{widetext}
A change of variables is needed to solve these coupled differential equations. We set $z=\ln{r}$ and re-write \eqref{delvr} and \eqref{delvtheta} as follows:
%\begin{widetext}
\begin{align}
	4\left(\partial_z^2-\frac{4+n^2}{4}\right)\delta v_{r,n}+3 in\left(\partial_z-\frac{5}{3}\right)\delta v_{\theta,n} &=h_n(z)\label{delvrz}\\
	\left(\partial_z^2 -(1+4n^2)\right)\delta v_{\theta,n}+3 in\left(\partial_z+\frac{5}{3}\right)\delta v_{r,n}
	 &=g_n(z)\label{delvtz}
\end{align}
%\end{widetext}
where
\begin{align}
h_n(z)&=- \frac{\zeta\Delta\mu}{\eta} e^{z}\left[\left(\partial_z+2\right)\delta\tilde{Q}_n(e^z) + in\delta q_n(e^z)\right]\\
g_n(z)&=-\frac{\zeta\Delta\mu}{\eta} e^{z}\left[\left(\partial_z+2\right)\delta q_n(e^z) - in \delta\tilde{Q}_n(e^z)\right]
\end{align}
which are functions we know exactly, in light of \eqref{delQn} and \eqref{delqn}. Acting with $\left(\partial_z-\frac{5}{3}\right)$ on \eqref{delvtz}, and using \eqref{delvrz}, we can eliminate $\delta v_{\theta,n}$ from these equations. Then, for modes $n\geq2$, the solution for $\delta v_{r,n}$ is of the form
\begin{widetext}
\begin{multline}
\delta v_{r,n} =\frac{-4+9n^2}{72n\eta} \Bigg[ \frac{C_1}{r^{n+1}}+\frac{C_2}{r^{n-1}}+C_3 r^{n-1}+C_4 r^{n+1}-\frac{1}{(n+1)r^{n+1}}\int_{R_0}^rx^{n}f_n(x)dx\\
    +\frac{1}{(n-1)r^{n-1}}\int_{R_0}^rx^{n-2}f_n(x)dx-\frac{r^{n-1}}{(n-1)}\int_{R_0}^r\frac{f_n(x)}{x^{n}}dx + \frac{r^{n+1}}{(n+1)}\int_{R_0}^r\frac{f_n(x)}{x^{n+2}}dx\Bigg]\label{vrsol}
\end{multline}

\end{widetext}
$f_n(x) =f_n(z)$ where $x\to e^{z}$ and \\
$f_n(z) = \frac{\left(\partial_z^2 -(1+4n^2)\right)}{4}h_n(z)-\frac{3in}{4}\left(\partial_z-\frac{5}{3}\right)g_n(z)$ 
which involves the exact solutions of $\delta\tilde{Q}_n$ and $\delta q_n$, given in \eqref{delQn} and \eqref{delqn}.
For the expression to not blow up at $r\to\infty$, we must enforce the following equalities:
\begin{align}
	 C_3 &=\frac{1}{ (n-1)}\int_{R_0}^{\infty}\frac{f_n(x)}{x^{n}}dx\\
    C_4 &= -\frac{1}{(n+1)}\int_{R_0}^{\infty}\frac{f_n(x)}{x^{n+2}}dx
\end{align}
Now, to get the expression for $\delta v_{\theta,n}$, we use a simple trick. We apply $\left(\partial_z+\frac{5}{3}\right)$ on \eqref{delvrz}, which allows us to derive $\partial_z^2 \delta v_{\theta,n}$. We substitute this in \eqref{delvtz} to get an expression of $\delta v_{\theta,n}$ in terms of $\delta v_{r,n}$, $\delta \tilde{Q}_n$ and $\delta q_n$. All of this leaves us with two undetermined constants $C_1$ and $C_2$ which have to be evaluated from the stress boundary conditions which set normal component of stress tensor at $r=R(\theta)$ to 0. This implies
\begin{align}
	(\sigma_{rr}n_r+\sigma_{r\theta}n_{\theta})\Bigr|_{r=R(\theta)}&=0\\
	(\sigma_{\theta r}n_r+\sigma_{\theta\theta}n_{\theta})\Bigr|_{r=R(\theta)}&=0
\end{align}
where $\hat{n}$ is the normal vector at the deformed ring $R(\theta)$. We can evaluate $\hat{n}$ to find:
\begin{equation}
\hat{n}(\theta)=-\hat{r}+\frac{1}{R_0}\frac{d\delta R}{d\theta}\hat{\theta}\label{normal}
\end{equation}
where terms only up to first order in the perturbation have been kept. Hence, we can use \eqref{normal} in the stress boundary conditions to obtain the following boundary conditions:
\begin{align}
	\delta\sigma_{rr}(R_0)&=-\partial_r\sigma^0_{rr}(R_0)\delta R(\theta)\\
	\delta\sigma_{\theta r}(R_0) &= \frac{1}{R_0}\frac{d\delta R}{d\theta}\sigma^0_{\theta\theta}
\end{align}
which implies, for the Fourier components,
\begin{align}
	\delta\sigma_{rr,n}(R_0)&=-\partial_r\sigma^0_{rr}(R_0)\delta R_n\\
	\delta\sigma_{\theta r,n}(R_0) &= \frac{in}{R_0}\delta R_n \sigma^0_{\theta\theta}
\end{align}
Using the above boundary conditions, $C_1$ and $C_2$ can be calculated, thereby producing the full exact solution expressed in \eqref{vrsol}.

For the modes $ n=0,1 $, the above four solutions are not independent hence these modes require a separate analysis. In fact, for $ n=1 $, a non-perturbative argument based on translation symmetry (as explained in \cite{alert_thesis}) shows that the mode should be exactly marginal i.e. $ \omega_1=0 $. For $ n=0 $, however, we need to do some more work. The form of the solution for n=0 is:
\begin{equation}
\delta v_{r,0} = C_1 r+\frac{C_2}{r}+r\int_{R_0}^r \frac{g_0(x)}{x^2}dx -\frac{1}{r}\int_{R_0}^r g_0(x)dx
\end{equation}
where $g_0(x)=-\dfrac{\zeta\Delta\mu}{4\eta }(\partial_r+\frac{2}{r})\delta\tilde{Q}_0$. To prevent $\delta v_{r,0}$ from blowing up at $r\to\infty$, we must enforce
\begin{equation}
C_1=-\int_{R_0}^{\infty}\frac{g_0(x)}{x^2}dx
\end{equation}
To evaluate $C_2$, we use the stress-free boundary conditions at $r=R(\theta)$ as discussed already. This gives us the full solution for $\delta v_{r,0}$. The rest of the analysis can be continued as for other modes.

\textit{Kinematic boundary condition} ---
We define the ring velocity $V(\theta) = (v_r \hat{r}+v_\theta\hat{\theta})\cdot\hat{n}\Bigr|_{R(\theta)}$, where $R(\theta)=R_0+\delta R(\theta)$. Splitting $\vec{v}$ into zeroth order solution and perturbation, we find:
\begin{align}
	V(\theta)&=\bigm(v^0_r(r) \hat{r}+ \delta v_r(r,\theta) \hat{r} + \delta v_\theta(r,\theta) \hat{\theta} \bigm)\cdot\hat{n}\Bigr|_{R(\theta)}\notag\\
	&=-v^0_r(R) - \delta v_r(R) - \frac{\delta v_\theta(R)}{R_0}\frac{d\delta R}{d\theta}\notag\\
	&\approx -\left(v_r^0(R_0)+\partial_r v_r^0(R_0)\delta R(\theta)\right)-\delta v_r(R_0,\theta)\notag\\
\therefore	\delta V(\theta)&\equiv V(\theta)-(-v_r^0(R_0))\notag\\
	&=-\partial_r v_r^0(R_0)\delta R(\theta)-\delta v_r(R_0,\theta)
\end{align}
Converting to Fourier modes again, with $\delta V_n=\dfrac{1}{2\pi}\int_0^{2\pi}\delta V(\theta)e^{-in\theta}$, we have:
\begin{equation}
\delta V_n=-\partial_r v_r^0(R_0)\delta R_n-\delta v_{r,n}(R_0)
\end{equation}
We now need to set up a kinematic boundary condition. This relates $\delta V$ to $\delta R$ as:
\begin{equation}
\delta V_n=-\frac{d\delta R_n}{dt}\label{kinematic}
\end{equation}
which gives us a way to evaluate $\omega_n$ - the growth/decay rate of the nth Fourier mode. The time dependence of the Fourier modes can be put in as follows: 
\begin{equation}
\delta R(\theta,t)=\sum_{n=0}^{\infty} \delta R_n e^{in\theta+\omega_n t}
\end{equation}
Therefore, $\dfrac{d\delta R_n}{dt}=\omega_n R_n$. So we derive, from \eqref{kinematic},
\begin{equation}
\omega_n=\partial_r v_r^0(R_0)+\frac{\delta v_{r,n}(R_0)}{\delta R_n}
\end{equation}

\begin{figure}
	\centering
	\includegraphics[width=0.9\linewidth]{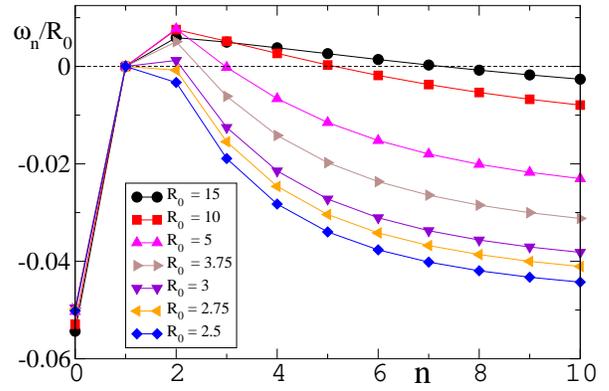}
	\caption{$\omega_n/R_0$ versus $ n $ for different values of $R_0$. The units of length and time are $ L_c\simeq 1\mu m$ and $ \frac{\eta}{\zeta\Delta\mu}\simeq 2sec$, respectively.}
	\label{fig:omega_n}
\end{figure}

We performed the calculation outlined above on Mathematica, choosing a set of parameter values that are experimentally relevant (given in main text). We have plotted $ \omega_n $ as a function of mode number $ n $. We find that the unstable modes are present in a small window for large values of the dimensionless radius $ R_0/L_c $. As the ring radius becomes smaller, the window of instability slowly diminishes in size and finally, at small enough radius, there are no unstable modes left. This parallels the experimental observation that asymmetrical modes are prominent when the contracting ring is large but they essentially disappear when the ring contracts to a small size, turning it into an increasingly perfect circle.

%\clearpage
%\bibliography{./local.bib}